\documentclass[a4paper,fleqn,usenatbib]{mnras}

\usepackage{newtxtext,newtxmath}

\usepackage[T1]{fontenc}
\usepackage{ae,aecompl}

\usepackage{graphicx}	
\usepackage{amsmath}	
\usepackage{amssymb}	
\usepackage{hyperref}
\usepackage{tabularx}
\usepackage{gensymb}
\hypersetup{pdfauthor={G. Halevi},pdftitle={r-Process Nucleosynthesis from Three-dimensional Jet-driven Core-Collapse Supernovae with Magnetic Misalignments},bookmarksnumbered=true}
\newcommand{\msun}{M$_\odot$}

\title[$r$-Process in Misaligned Jet-driven SNe]{$r$-Process Nucleosynthesis from Three-dimensional Jet-driven Core-Collapse Supernovae with Magnetic Misalignments}

\author[G. Halevi et al.]{
Goni Halevi$^{1,2}$\thanks{E-mail: \href{mailto:ghalevi@astro.princeton.edu}{ghalevi@astro.princeton.edu}} \&
Philipp M\"{o}sta$^{2}$\thanks{NASA Einstein Fellow}
\\
$^{1}$Department of Astrophysical Sciences, 4 Ivy Lane, Princeton University, Princeton, NJ 08544\\
$^{2}$Department of Astronomy, 501 Campbell Hall \#3411, University of California at Berkeley, Berkeley, CA 94720\\
}

\date{Accepted XXX. Received YYY; in original form ZZZ}

\pubyear{2018}

\begin{document}
\label{firstpage}
\pagerange{\pageref{firstpage}--\pageref{lastpage}}
\maketitle

\begin{abstract}
We investigate $r$-process nucleosynthesis in three-dimensional general relativistic magnetohydrodynamic simulations of jet-driven supernovae resulting from rapidly rotating, strongly magnetized core-collapse.  We explore the effect of misaligning the pre-collapse magnetic field with respect to the rotation axis by performing four simulations: one aligned model and models with 15$\degree$, 30$\degree$, and 45$\degree$ misalignments. The simulations we present employ a microphysical finite-temperature equation of state and a leakage scheme that captures the overall energetics and lepton number exchange due to post-bounce neutrino emission and absorption. We track the thermodynamic properties of the ejected material with Lagrangian tracer particles and analyse its composition with the nuclear reaction network \textsc{SkyNet}. By using different neutrino luminosities in post-processing the tracer data with \textsc{SkyNet}, we constrain the impact of uncertainties in neutrino luminosities. We find that, for the aligned model considered here, the use of an approximate leakage scheme results in neutrino luminosity uncertainties corresponding to a factor of 100-1000 uncertainty in the abundance of third peak $r$-process elements. Our results show that for misalignments of 30$\degree$ or less, $r$-process elements are robustly produced as long as neutrino luminosities are reasonably low ($\lesssim 5 \times 10^{52}$ erg s$^{-1}$). For a more extreme misalignment of 45$\degree$, we find the production of $r$-process elements beyond the second peak significantly reduced. We conclude that robust $r$-process nucleosynthesis in magnetorotational supernovae requires a progenitor stellar core with a large poloidal magnetic field component that is at least moderately (within $\sim30\degree$) aligned with the rotation axis.
\end{abstract}




\section{Introduction}

Rapid neutron capture ($r$-process) nucleosynthesis is required to explain the observed abundances of stable nuclei heavier than iron, yet the dominant astrophysical site at which it takes place has not been conclusively determined. While core-collapse supernovae (CCSNe) have long been considered a potential site for the production of heavy elements through $r$-process nucleosynthesis \citep[e.g.][]{1994ApJ...433..229W}, simulations of neutrino-driven CCSNe have failed to robustly produce the necessary conditions. The high event rates for CCSNe also make them less promising candidates than they were originally thought to be. Although early observations of old stars in the galactic halo suggested that $r$-process elements are continuously produced, more recent observations of stars in Reticulum II, an ultrafaint dwarf galaxy, instead favor the scenario of rare events with relatively high yields of $r$-process elements per event \citep{2016Natur.531..610J} to explain abundances in dwarf galaxies.

Due to these challenges in invoking CCSNe as dominant sites of $r$-process nucleosynthesis, mergers of double neutron star (NS-NS) and black hole (BH)-NS binaries have been heavily explored as alternative astrophysical sites \citep[e.g.][]{2010MNRAS.406.2650M,2013ApJ...773...78B,2014ApJ...789L..39W,2015MNRAS.448..541J}. Recently, LIGO detected the first gravitational wave signal produced by the coalescence of a NS-NS binary \citep{2017PhRvL.119p1101A}. This event (GW170817) was coincident with an electromagnetic counterpart \citep{2017ApJ...848L..12A} that was largely predicted by the kilonova models \citep{2017LRR....20....3M} of e.g. \citet{2013ApJ...775...18B} and \citet{2013ApJ...775..113T}. The agreement between models and observations suggests that a significant amount of $r$-process elements are produced in NS-NS mergers. However, there remain questions regarding the rates of such events and the relatively long delay time for binary coalescence compared to CCSNe. Additionally, recent observations \citep{2017arXiv171102121T,2017ApJ...850..179C} argue against NS-NS mergers as the exclusive producers of $r$-process elements. Given these remaining reservations, it is possible that CCSNe may at least be an important source of $r$-process elements in the early universe or in certain (e.g. low metallicity) environments. 

Magnetorotational (MR), or jet-driven, CCSNe \citep{1970AZh....47..813B,1970ApJ...161..541L,2002ApJ...568..807W,2014ApJ...785L..29M} appear to provide far more promising conditions for $r$-process nucleosynthesis than neutrino-driven CCSNe \citep{2012ApJ...750L..22W,2015ApJ...810..109N,2017ApJ...836L..21N}, ejecting $\sim 10^{-3}-10^{-2} M_\odot$ of $r$-process material. While this theoretical yield is comparable to predictions for NS-NS mergers \citep[e.g.][]{2013PhRvD..87b4001H}, the time-scales for MR CCSNe to occur are much shorter than for NS-NS binary coalescence due to the rapid evolution of massive stars. MR CCSNe are physically motivated as the engines driving hyperenergetic SNe from stripped-envelope progenitors, classified spectroscopically as SNe Type Ic-bl (H/He deficient with broad lines). Their event rate is in turn significantly lower than that of standard CCSNe, with MR CCSNe very roughly estimated to make up $\sim1$ per cent of all CCSNe based on SN populations, likely progenitors, and predictions of optical signatures for different SN engines.

The progenitor star for a jet-driven explosion must have rapid ($P_0 \simeq \mathcal{O}(1)\mathrm{s}$) iron core rotation \citep{2006ApJS..164..130O,2007ApJ...664..416B,2014ApJ...785L..29M} so that a ms-protoneutron star (PNS) forms after collapse. A magnetar-strength toroidal magnetic field component can be achieved through a combination of flux-compression from a highly magnetized progenitor core, amplification by the magnetorotational instability \citep[MRI,][]{1991ApJ...376..214B,2003ApJ...584..954A,2009A&A...498..241O,2016MNRAS.460.3316R}, and dynamo action \citep{2015Natur.528..376M} soon after core bounce. The magnetic field then funnels accreted material into an MHD jet \citep{1976ApJ...204..869M,2002ApJ...568..807W,2007ApJ...664..416B,2014ApJ...785L..29M} which drives the explosion. The differences in the progenitors and corresponding dynamics of jet-driven CCSNe compared to neutrino-driven CCSNe result in different outflow geometries and thermodynamic properties. In particular, the highly-magnetized, low density, neutron-rich (electron fraction $Y_e \simeq 0.1-0.3$) ejecta of jet-driven CCSNe is far more conducive to the onset of $r$-process nucleosynthesis than that of typical neutrino-driven CCSNe.

Past analyses of heavy element production in jet-driven explosions have used both axisymmetric (2D) and three-dimensional (3D) magnetohydrodynamic (MHD) simulations. In a 3D simulation of the collapse of a rare progenitor configuration with rapid rotation and high magnetization, \citet{2012ApJ...750L..22W} obtained a bipolar jetted explosion which robustly produced $r$-process elements in agreement with the solar abundance pattern, even when including the effects of neutrinos. Their assumed poloidal magnetic field ($B_\mathrm{pol} = 5 \times 10^{12}\mathrm{G}$) resulted in a jet powerful enough to resist disruption by the $m=1$ kink instability \citep{2014ApJ...785L..29M}. \citet{2015ApJ...810..109N} studied $r$-process nucleosynthesis for a range of CCSN simulations in axisymmetry, with an emphasis on MR SNe. They classified their explosion models into prompt and delayed magnetic jets and found that $r$-process occurs robustly for the former but is suppressed past the second peak (mass number $A\sim 130$) for the latter, concluding that heavy $r$-process elements may be formed in CCSNe if their magnetic fields are sufficiently strong. \citet{moestarobertshalevi2017} studied $r$-process nucleosynthesis for a suite of MR CCSN general relativistic (GR) MHD simulations, including 3D models with initial field strengths of $10^{13}$ and $10^{12}\mathrm{G}$. They concluded that the dynamics of the simulations strongly affect nucleosynthetic yields, and that only cores with large pre-collapse fields ($B\sim 10^{13}\mathrm{G}$) can result in MR CCSNe that robustly produce $r$-process elements.

In this work, we investigate $r$-process nucleosynthesis in full 3D dynamical-spacetime GRMHD simulations of MR CCSNe with initial field strengths of $B=10^{13}\mathrm{G}$. We study, for the first time, the effect of misalignment between the initial rotation and magnetic axes on $r$-process nucleosynthesis in MR CCSNe. Exploring misalignment probes the importance of the magnetic field configuration on the pre-collapse core in determining the explosion dynamics. We include one model in which the axes are aligned, which is identical to model B13 of \citet{moestarobertshalevi2017}. As in that work, the nucleosynthetic signature for each model is calculated by post-processing Lagrangian tracer particles from our simulations with the open-source nuclear reaction network \textsc{SkyNet} \citep{2017ascl.soft10005L}. We also explore the impact of neutrinos on $r$-process production by systematically varying the neutrino luminosities applied in the nuclear reaction network calculation. The simulations themselves use an approximate leakage scheme for neutrino transport so that the neutrino luminosities they output are uncertain to a factor of $\sim 2$.

We find that for neutrino luminosities on the lower end ($\lesssim 5 \times 10^{52}~\mathrm{erg}~\mathrm{s}^{-1}$) and moderate ($\lesssim 30$$\degree$) misalignments between the magnetic and rotation axes, $r$-process elements, including the third peak, are robustly produced. Both higher neutrino luminosities and significant (45$\degree$) misalignment of the two axes appreciably limit the production of the heaviest (second peak and beyond) $r$-process elements. The explosion dynamics are affected by initial misalignment, with all misaligned models showing noticeable deviations from the completely aligned case. However, the outflows in models with relatively mild (15$\degree$ and 30$\degree$) misalignments generally remain bipolar, while 45$\degree$ misalignment results in non-jetted, more isotropic ejecta. Our main conclusion is that we require a strong, poloidal magnetic field that is not severely misaligned with the rotation axis on the pre-collapse core in order to robustly generate $r$-process elements from a jet-driven CCSNe. This is in agreement with the results of \citet{moestarobertshalevi2017}, but extends their findings to more general magnetic field configurations.

The organization of this paper is as follows. In Sec. \ref{sec:methods}, we introduce the methods and setup used for our simulations and post-processing nuclear reaction network. In Sec. \ref{sec:results}, we discuss our results, including the MHD dynamics, ejecta properties, and nucleosynthesis yields. We then present a discussion and interpretation of these results in Sec. \ref{sec:disc} and end with a summary and our main conclusions in Sec. \ref{sec:conc}.

\begin{figure*}
    \includegraphics[width=0.49\textwidth]{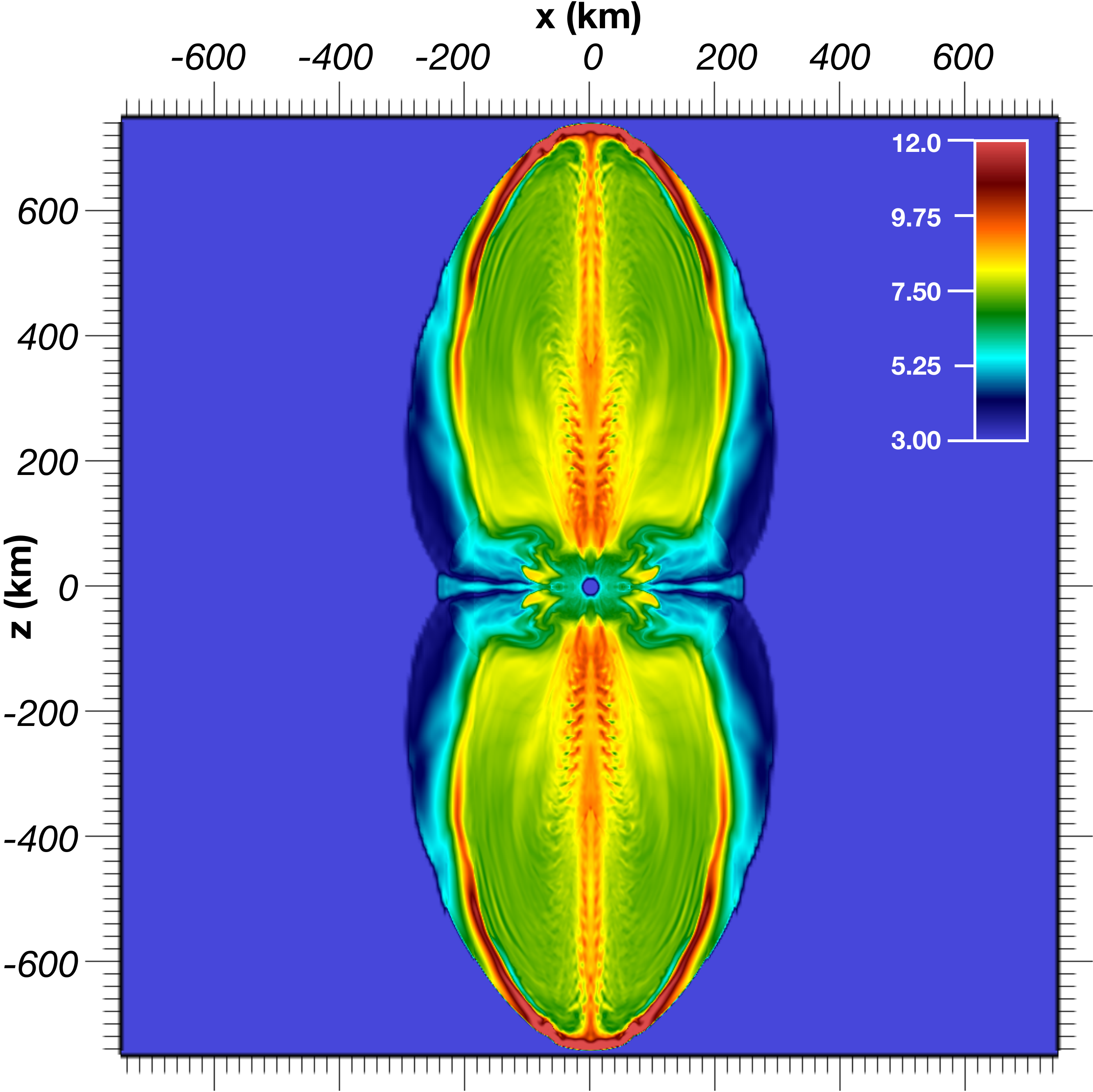}
    \includegraphics[width=0.49\textwidth]{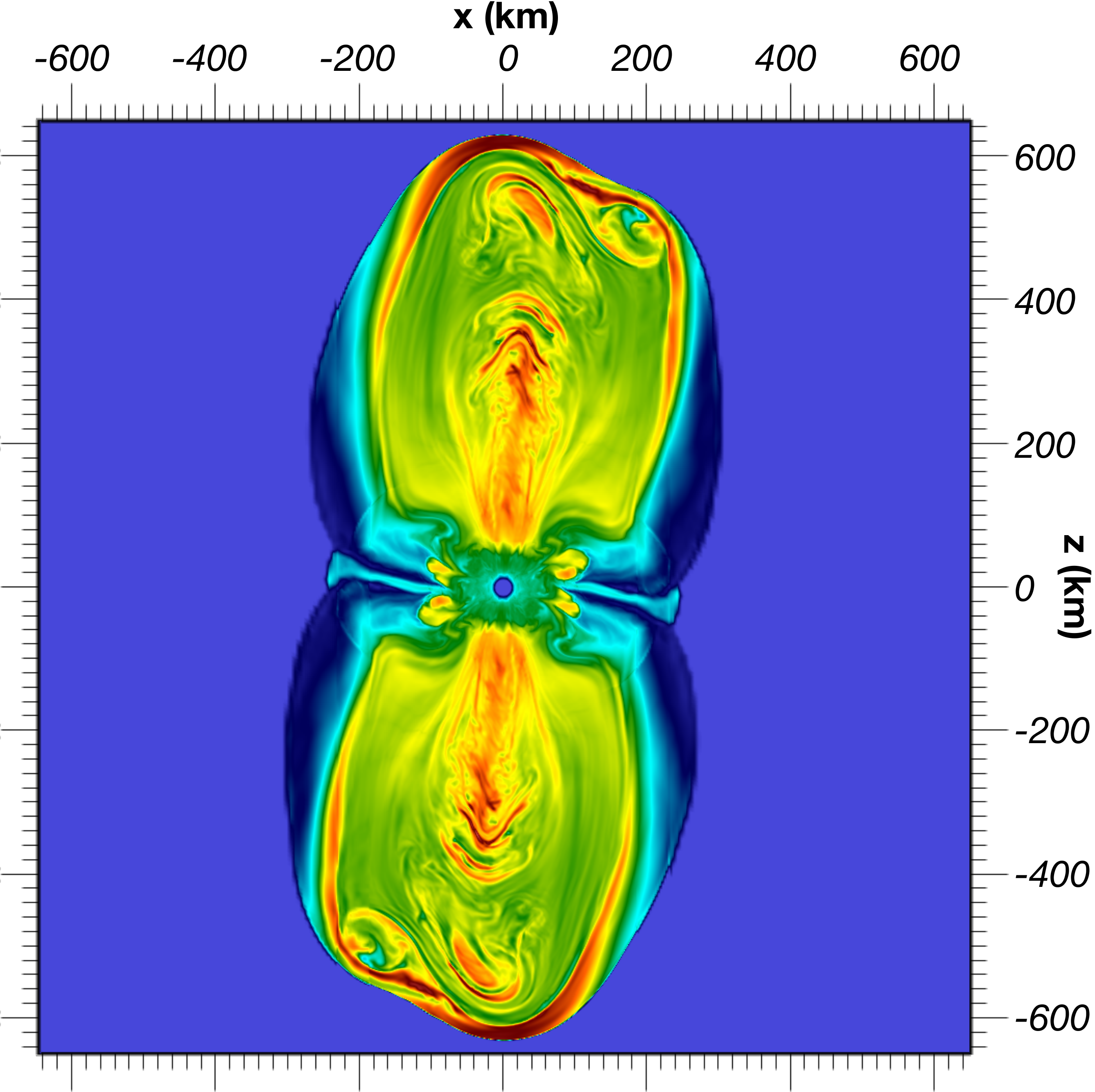}
    \includegraphics[width=0.49\textwidth]{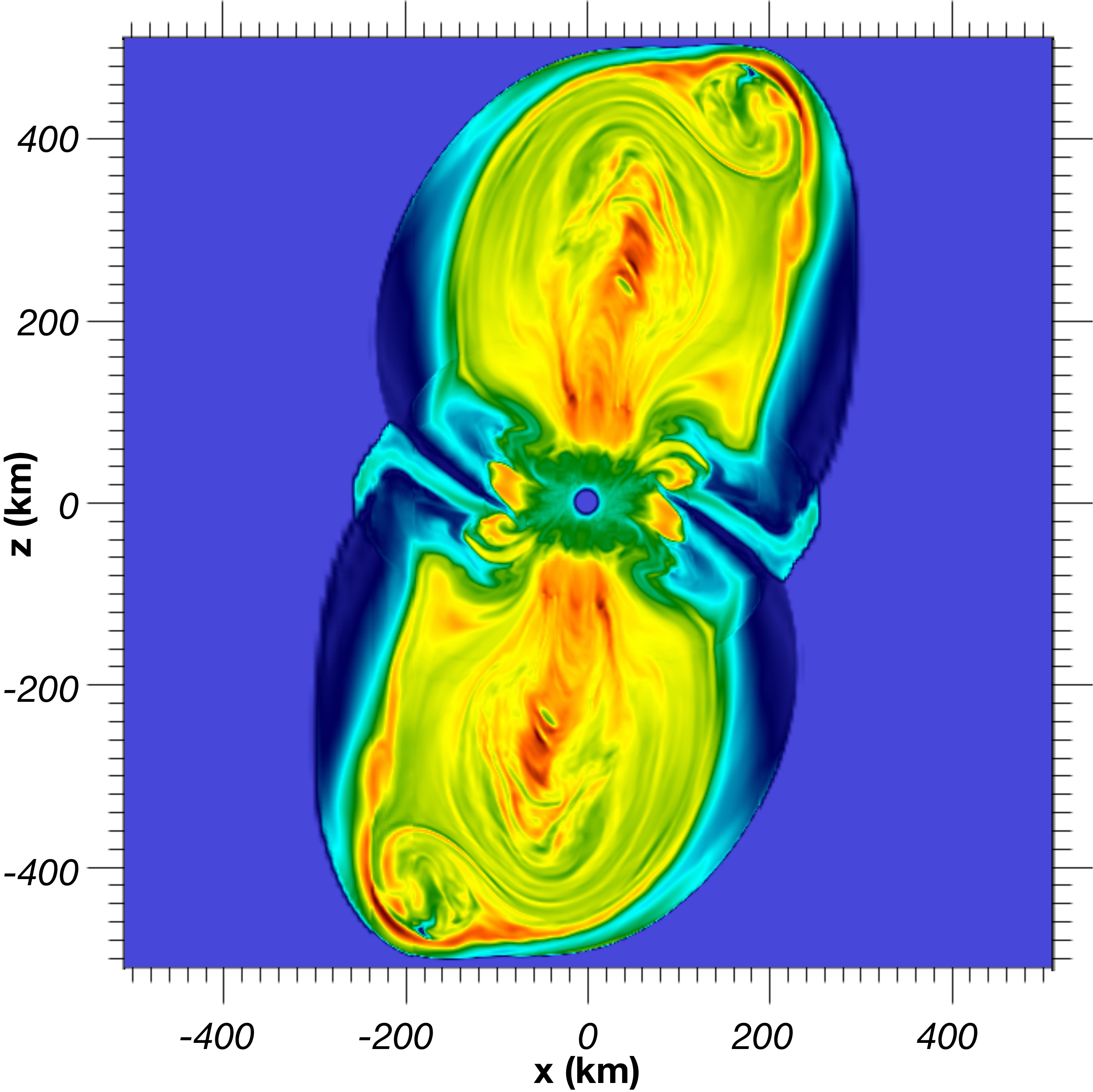}
    \includegraphics[width=0.49\textwidth]{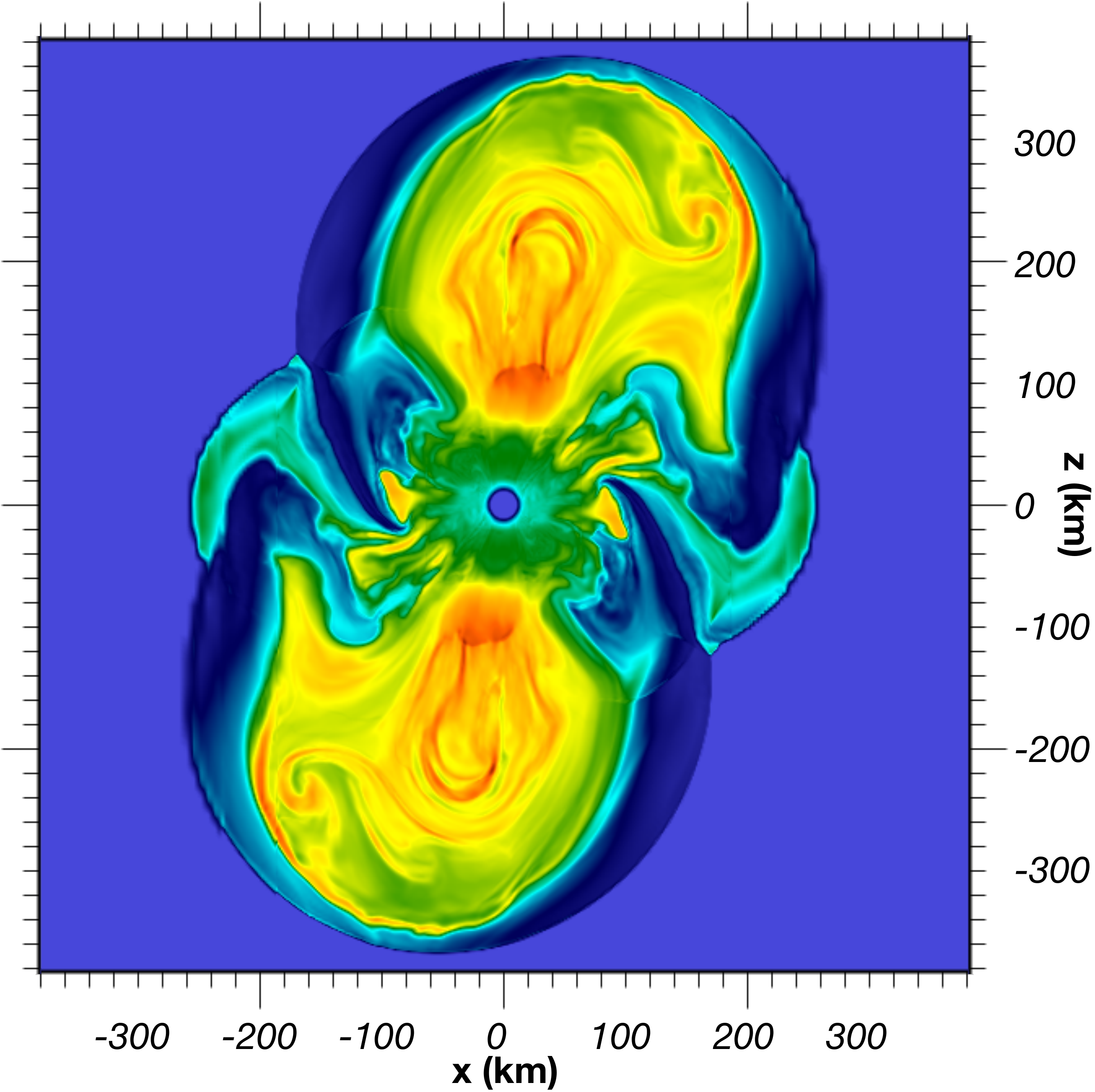}
    \caption{Two-dimensional slices through the rotation axes of volume renderings for the simulations showing specific entropy $s$ at 20 ms after core bounce. The panels show the four cases of: alignment (top left), 15$\degree$ misalignment (top right), 30$\degree$ misalignment (bottom left), and 45$\degree$ misalignment (bottom right) between pre-collapse magnetic and rotation axes. The colourbar is the same for all panels, and is shown in units of $k_b~\mathrm{baryon}^{-1}$. The physical scale differs between panels and is indicated for each.}
    \label{fig:ent_xz}
\end{figure*}

\begin{figure*}
    \includegraphics[width=0.49\textwidth]{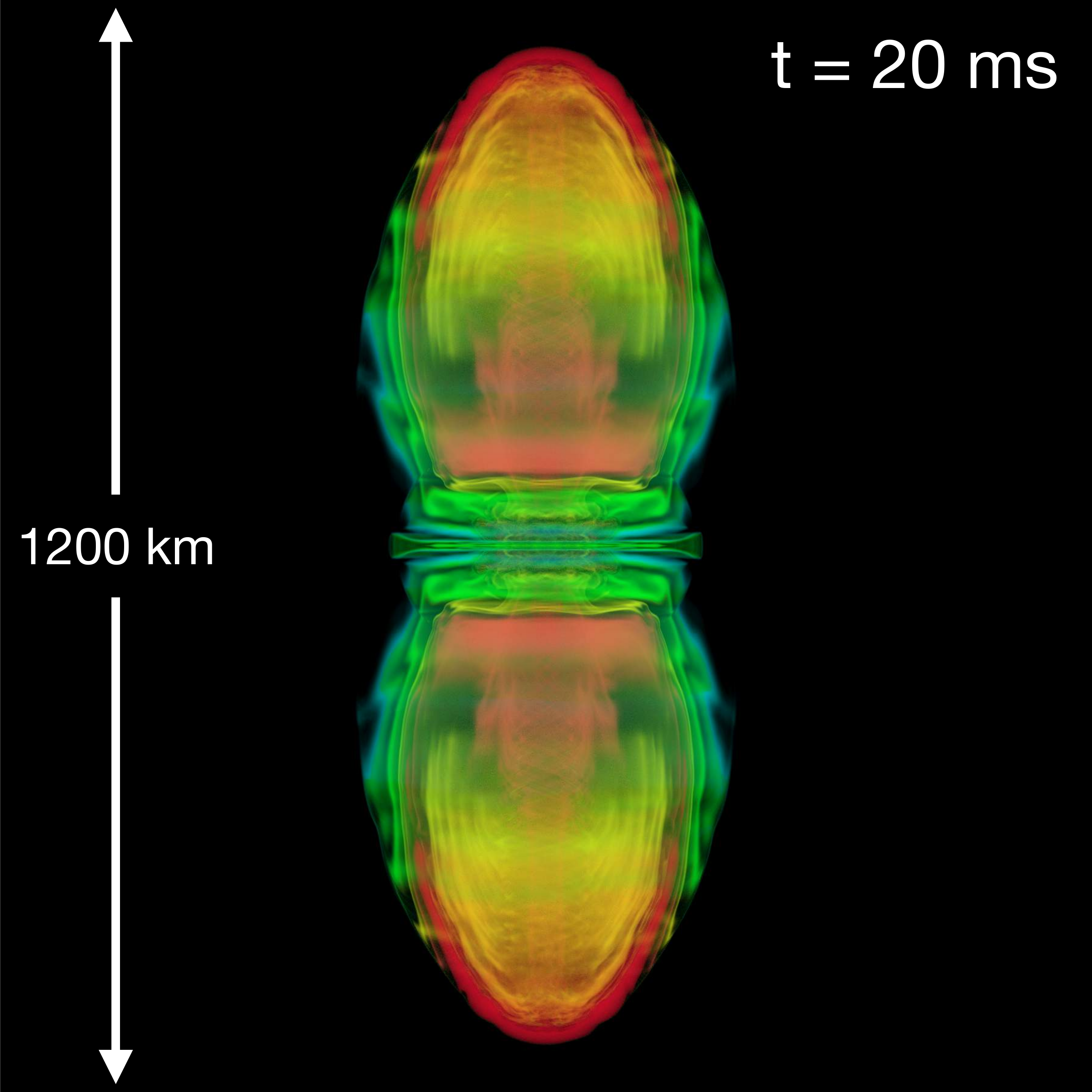}
    \includegraphics[width=0.49\textwidth]{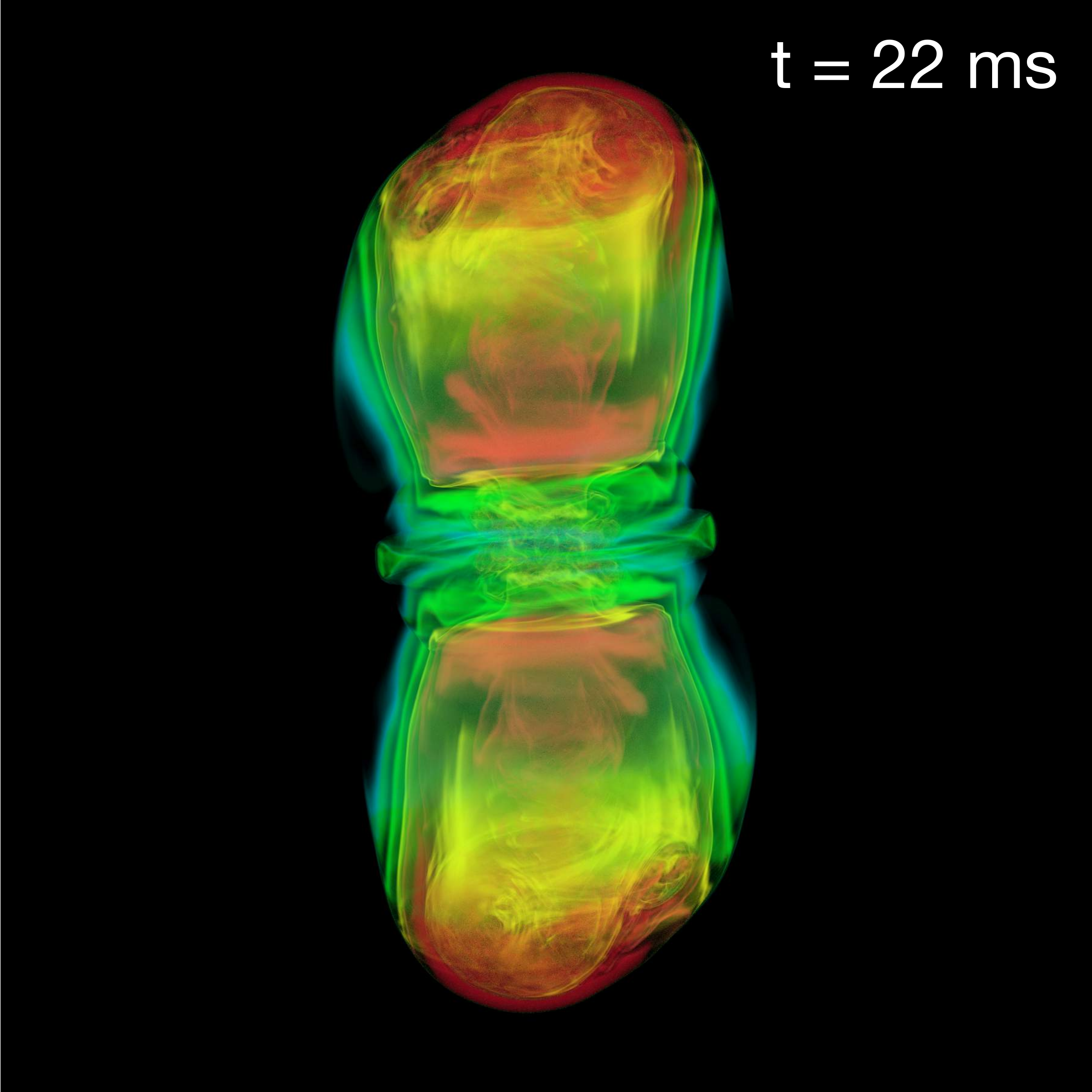}
    \includegraphics[width=0.49\textwidth]{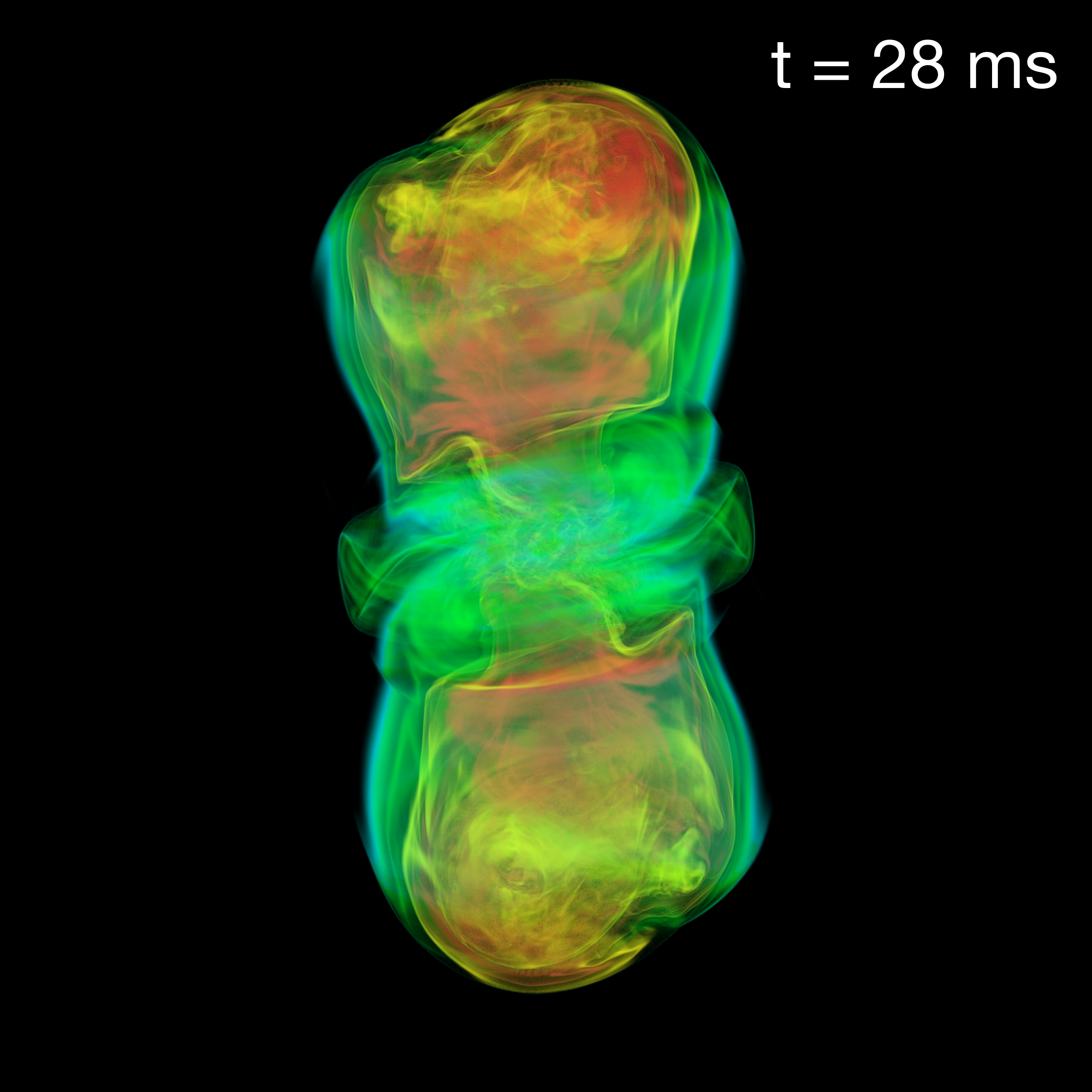}
    \includegraphics[width=0.49\textwidth]{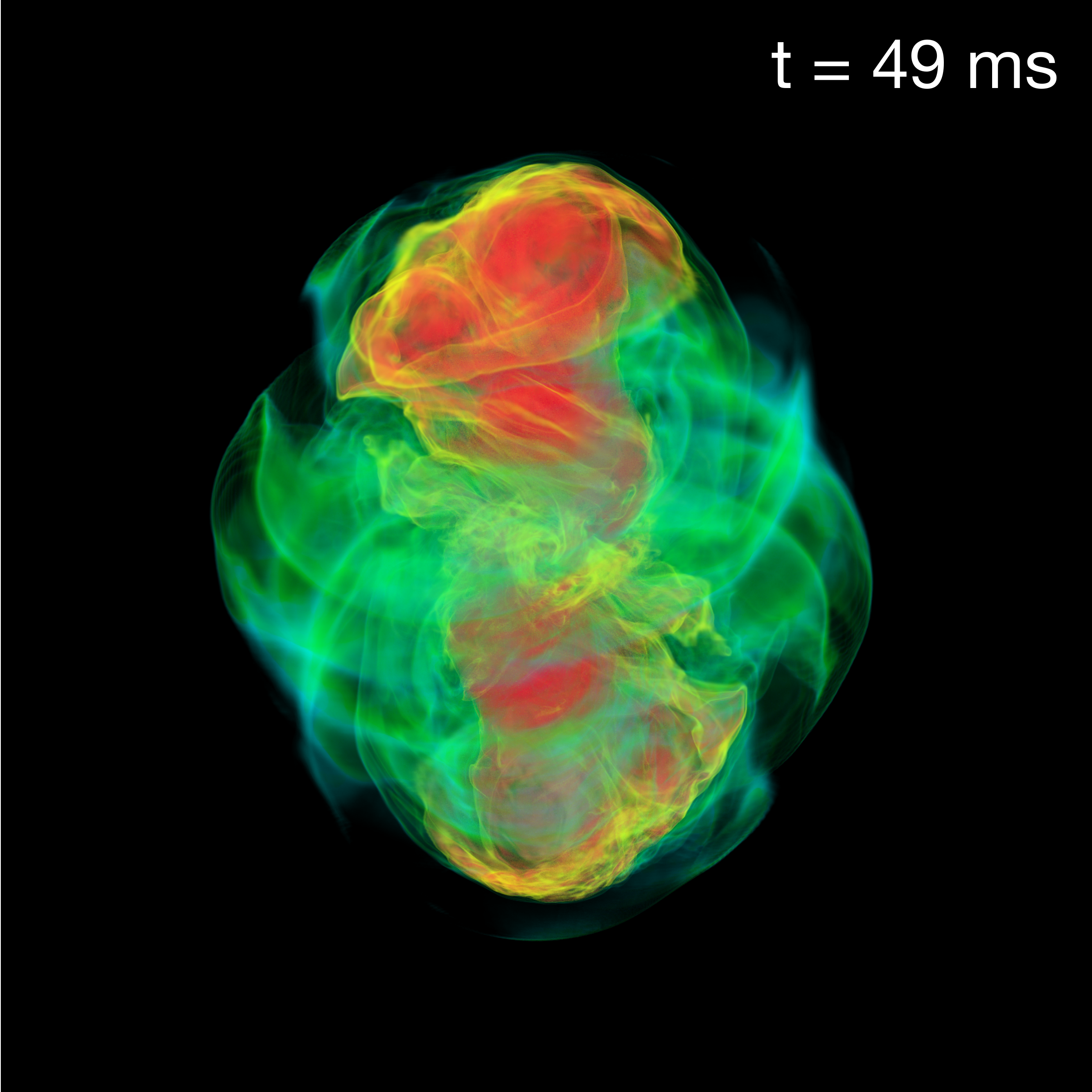}
    \caption{Volume renderings of specific entropy for the four simulations, with the panels corresponding to those shown in Fig. \ref{fig:ent_xz} at the post-bounce times indicated. We select the colourmap such that blue corresponds to $s = 3.7$, cyan to $s = 4.8$ (indicating the shock surface), green to $s = 5.8$, yellow to $s = 7.4$, and red to $s=10$, in units of $k_b~\mathrm{baryon}^{-1}$. The physical size scale, 1200 km x 1200 km, is identical for all four panels. Full movies showing the time evolution for each of these four volume renderings are available at \url{http://www.astro.princeton.edu/~ghalevi/mrsne_movies}.}
    \label{fig:volren}
\end{figure*}

\section{Methods and Setup}
\label{sec:methods}

\subsection{Simulations}
\label{sec:sims} 

We perform three-dimensional (3D) ideal magnetohydrodynamic (MHD) simulations with fully-dynamical general relativity (GR) and adaptive mesh refinement (AMR) using the open-source \texttt{EinsteinToolkit}~\citep{2014CQGra..31a5005M,2012CQGra..29k5001L}. We implement GRMHD using finite-volume with WEN05 reconstruction~\citep{2013PhRvD..87f4023R,2007MNRAS.379..469T} and use the HLLE Reimann solver~\citep{1988stw..proc..671E} and constrained transport~\citep{2000JCoPh.161..605T} to maintain a divergence-free magnetic field.

We employ the $K_0 = 220\,\mathrm{MeV}$ variant of the equation of state of \citet{1991NuPhA.535..331L} and the approximate neutrino leakage and heating prescriptions as described in \citet{2010CQGra..27k4103O} and \citet{2012PhRvD..86b4026O}. Prior to collapse, we cover the inner $\sim5700\,\mathrm{km}$ of the core with four AMR levels, adding five more levels during collapse. After core bounce, the protoneutron star (PNS) is covered with a grid resolution of $\sim370\,\mathrm{m}$ and the AMR structures consist of boxes with extent [$5674.0\,\mathrm{km}$, $3026.1\,\mathrm{km}$, $2435.1\,\mathrm{km}$, $1560.3\,\mathrm{km}$, $283.7\,\mathrm{km}$, $212.8\,\mathrm{km}$, $144.8\,\mathrm{km}$, $59.1\,\mathrm{km}$, $17.7\,\mathrm{km}$].  The coarsest resolution is $h = 94.6\, \mathrm{km}$ and refined meshes differ in resolution by factors of 2. We use adaptive shock tracking to ensure that the shocked region is always contained on the mesh refinement box with resolution $h = 1.48\, \mathrm{km}$.

We simulate the explosions of four highly magnetized and rapidly rotating stellar cores. The pre-collapse models are all initialized by adding rotation and a magnetic field to the 25\msun~(at zero-age-main-sequence) progenitor model (E25) of \citet{2000ApJ...528..368H}. We employ the rotation profile given by eqn. (1) of \citet{2011ApJ...743...30T} and used in~\citet{2014ApJ...785L..29M,moestarobertshalevi2017},
\begin{equation}
    \Omega(x,z) = \Omega_0 \frac{x_0^2}{x^2+x_0^2}\,\frac{z_0^4}{z^4+z_0^4}~,
\end{equation}
with $\Omega_0 = 2.8\, \mathrm{rad}\, \mathrm{s}^{-1}$, $x_0 = 500\,\mathrm{km}$, and $z_0 = 2000\,\mathrm{km}$. $\Omega_0$ sets the initial central angular velocity, while $x_0$ and $z_0$ set the fall-off in cylindrical radius and vertical position, respectively.

The magnetic field setup is a modified poloidal field which remains nearly constant in magnitude within a radius $r_0$ and dipolar at $r>r_0$. For the case of alignment with the rotation axis, the vector potential that defines this field setup is
\begin{subequations}
\begin{align}
   A_r &= A_\theta = 0, \\
   A_\phi &= \frac{B_0}{2} \frac{r_0^3}{r^3 + r_0^3}\, r \sin \theta
\end{align}
\end{subequations}
where we take an inner radius of $r_0 = 1000\, \mathrm{km}$ and an initial field strength of $B_0 = 10^{13}~\mathrm{G}$ for all models. The four models differ by the angle between the pre-collapse rotation and magnetic axes; we run simulations with initial misalignments of 0$\degree$, 15$\degree$, 30$\degree$, and 45$\degree$. Every model has its rotation axis along $\hat{z}$, but for the misaligned models, the scalar potential that sets the $B$-field is adjusted so that the magnetic axis is rotated away from $\hat{z}$. The modified magnetic field for a rotation angle $\alpha$ is derived from the vector potential:
\begin{subequations}
\begin{align}
	A_r &= 0, \\
	A_\theta &= -\frac{B_0}{2} \frac{r_0^3}{r^3 + r_0^3}\,r\sin\phi \sin \alpha , \\
	A_\phi &= \frac{B_0}{2} \frac{r_0^3}{r^3 + r_0^3}\, \left(r \sin \theta\cos \alpha - r\cos\theta \cos \phi \sin \alpha \right).
\end{align}
\end{subequations}

\subsection{Tracer particles and post-processing}
\label{sec:post}

\begin{figure*}
    \includegraphics[width=0.246\textwidth]{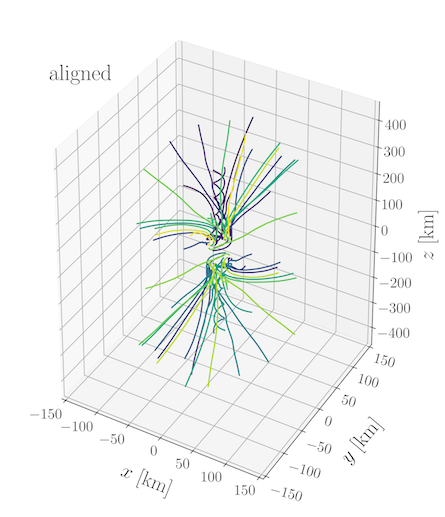}
    \includegraphics[width=0.246\textwidth]{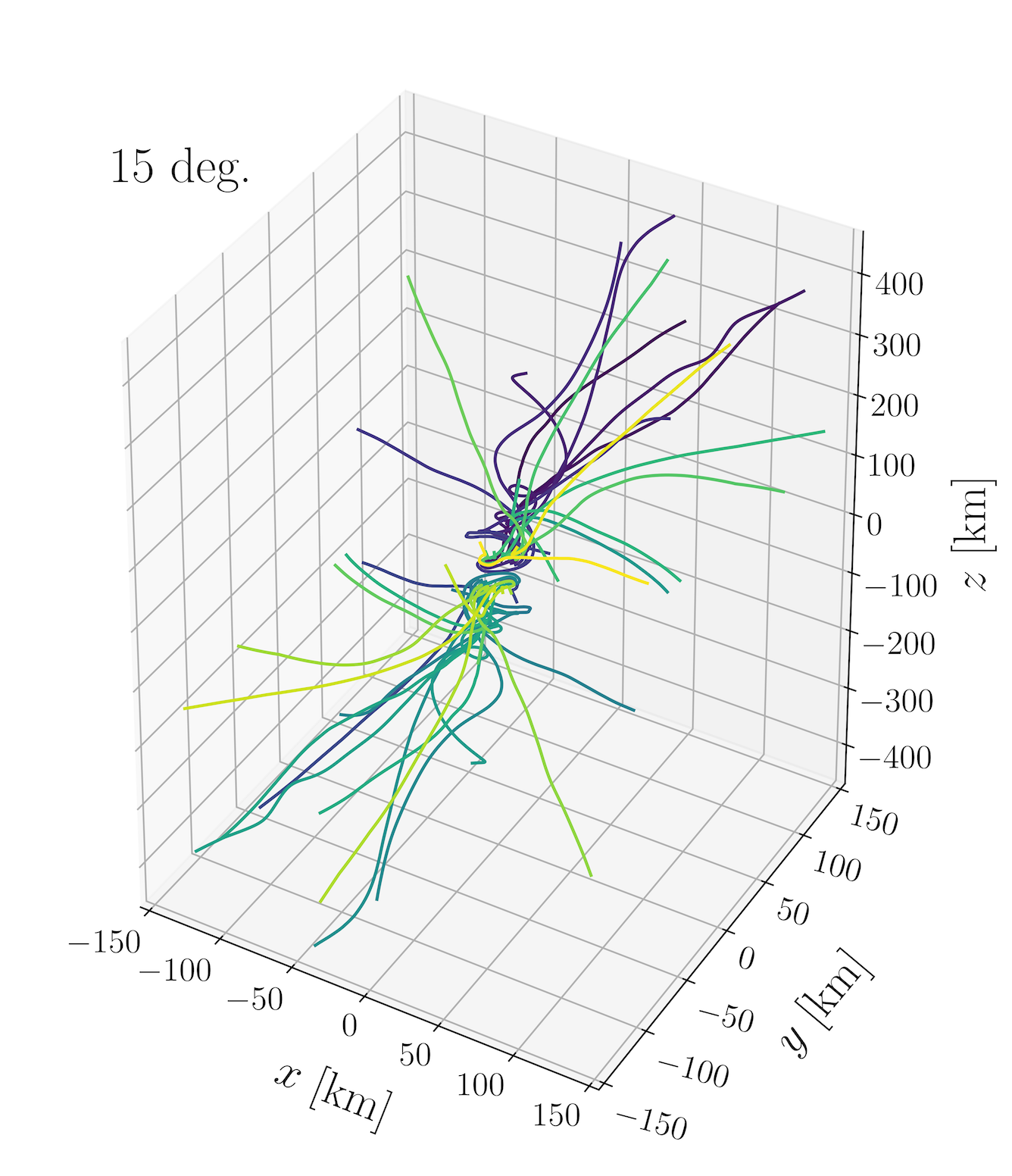}
    \includegraphics[width=0.246\textwidth]{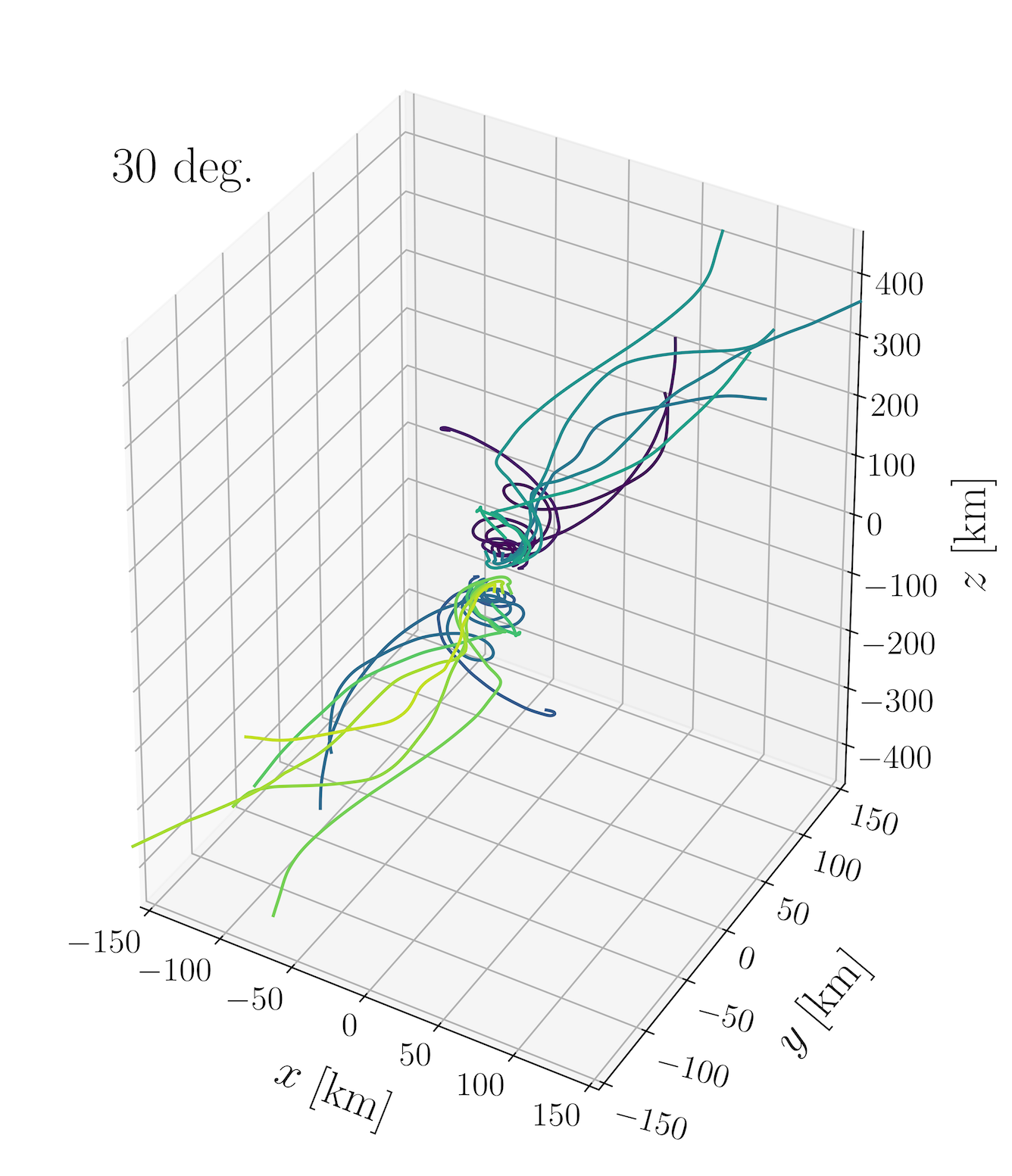}
    \includegraphics[width=0.246\textwidth]{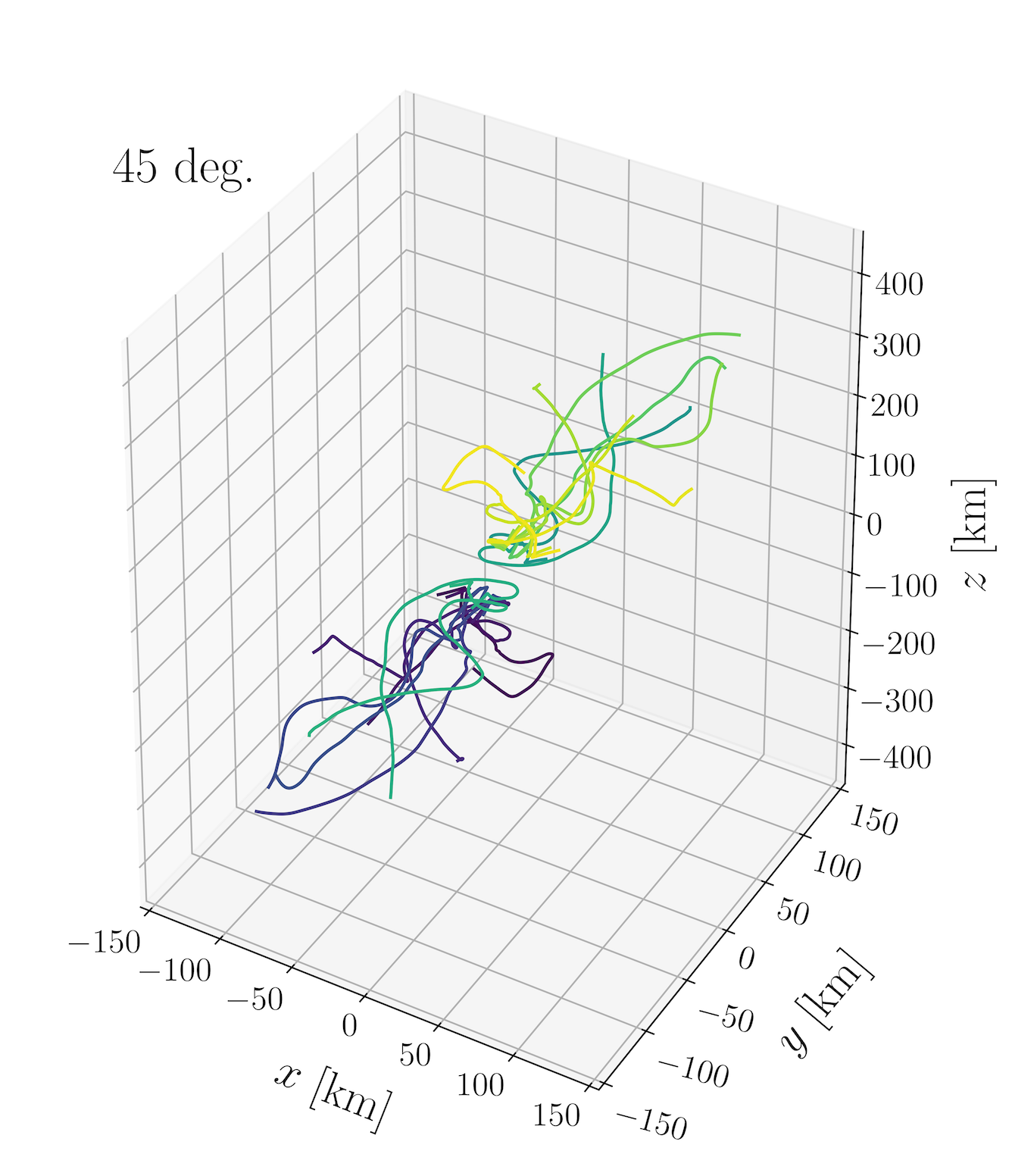}
    \caption{Spatial trajectories in 3D shown for a subset of individual tracer particles in each of the four simulations. Color is used to help distinguish between tracer particles and does not correspond to any particular property.}
    \label{fig:tracers3d}
\end{figure*}

In order to calculate nucleosynthesis yields, we record and extract thermodynamic properties of the ejected material via Lagrangian tracer particles. The particles are initialized at core bounce and advect with the hydrodynamic flows as the explosion proceeds. Data from each tracer particle as a function of time are obtained by interpolating from the 3D simulation grid, allowing us to record the thermodynamic conditions and neutrino luminosities they experience. At core bounce, we initialize 20,000 tracers uniformly in space so that they span a range of $\sim 30-1000~\mathrm{km}$ radially and represent regions of constant volume. We assign a mass to each tracer particle by taking into account the density at its initial location and the volume it occupies. In determining the mass ejected by the explosions, and therefore the nucleosynthetic abundance patterns of the ejecta, we consider only those particles which become dynamically unbound. We do not investigate convergence for the tracer particle resolution due to the high computational cost of running these simulations. However, because this work focuses on abundance patterns and comparisons between models rather than absolute measures of ejecta masses, we do not expect tracer particle resolution effects to modify our main results. Still, we recognize that a particle convergence study would be useful and plan to investigate particle resolution in the future.

Data from the tracer particles are read in and post-processed with the nuclear reaction network \textsc{SkyNet} \citep{2017ascl.soft10005L}, which includes isotopes up to $^{337}$Cn. We take forward strong rates from the JINA REACLIB database \citep{JINAreaclib} and compute the inverse rates under the assumption of detailed balance. Weak rates are taken from \citet{1982ApJS...48..279F}, \citet{1994ADNDT..56..231O}, \citet{2000NuPhA.673..481L} or also from REACLIB. We also utilize nuclear masses and partition functions from REACLIB. The network evolves the temperature by computing source terms due to individual nuclear reactions as well as neutrino interactions. \textsc{SkyNet} allows us to set the neutrino luminosities and energies to constant values in post-processing, which we take advantage of in this work by varying the neutrino luminosities. In all cases, we set the neutrino luminosity to be equal for electron neutrinos and electron antineutrinos (e.g. $L_\nu = L_{\nu_\mathrm{e}} = L_{\bar{\nu}_\mathrm{e}}$). We keep average neutrino energies at constant values of 12 and 15 MeV for the electron neutrinos and antineutrinos, respectively, throughout the network calculations.

In the remainder of this work, we present results calculated using constant neutrino luminosities in post-processing, ranging from neglecting neutrinos entirely ($L_\nu = 0$) to artifically high neutrino luminosities ($L_\nu = 10^{53}~\mathrm{erg}~\mathrm{s}^{-1}$), as well as results obtained by extracting neutrino luminosities from the tracer particle data. Our simulations employ a leakage scheme that captures the overall energetics and lepton number exchange due to postbounce neutrino emission and absorption, however this approximation produces small uncertainties in neutrino luminosities. Realistic neutrino luminosities likely differ by up to a factor of $\sim 2$ from those extracted by the tracer particles.

For each tracer particle, we take nuclear statistical equilibrium (NSE) as the starting point and begin the network calculations as soon as the temperature falls below $25~\mathrm{GK}$. However, our simulations run for tens of ms after core bounce, which is insufficient time to reach this temperature. Instead, we extrapolate particle data from the end of the simulation until the start of the network under the assumption of homologous expansion. This assumption is valid by the time we stop running our simulations, as seen in Fig. \ref{fig:traj}. The extrapolated conditions at the time when the temperature drops to $25~\mathrm{GK}$ serve as the initial conditions for the network calculation, which then proceeds for $10^9~\mathrm{s}$. At this time, stable abundance patterns (ejected mass as a function of mass number $A$) are generated.

\section{Results}
\label{sec:results}

\subsection{MHD Dynamics} \label{sec:MHDdyn}

For all four models, the collapse phase proceeds identically. Core bounce consistently occurs $\sim 440$ ms after the start of the simulation, differing by less than 2 ms between the four models. In all cases, the magnetic fields are amplified by compression and rotational winding during collapse, resulting in poloidal and toroidal field strengths of $\sim 10^{16}\mathrm{G}$.

For the aligned model, the shock propagates roughly spherically immediately after core bounce and does not stall at any point. It instead transitions into a mildly relativistic ($v\sim 0.1c$), jetted explosion along the rotation axis. This MHD jet, powered by magnetic pressure and stresses, is stable against the $m=1$ MHD kink instability due to the large poloidal component of the $B$-field. A slight $m=0$ deformation occurs (seen as a spiral structure in the top left panel of Fig. \ref{fig:volren}), but is too mild to distrupt the jet. The three misaligned models also produce shocks that continue to propagate without stalling, driving highly anisotropic explosions, but with differences in the jet collimation, propagations speeds, and overall geometries.

Snapshots of our simulations 20 ms after core bounce, coloured by specific entropy, are shown in Fig.~\ref{fig:ent_xz}. Each snapshot is a 2D slice through the full 3D simulation, taken along the rotation ($z$) axis. The aligned case (top left) shows a strong bipolar jet with high-entropy regions along the polar axis and in the outer lobes. The cases of 15$\degree$ (top right) and 30$\degree$ (bottom left) misalignments show twisted jets with high-entropy bubbles and more complex structure, while the case of 45$\degree$ misalignment (bottom right) shows much more spherical outflow and disordered high-entropy structures.

The differences in the dynamics between the four simulations are also apparent in 3D snapshots, shown in Fig. \ref{fig:volren}. These volume renderings, again coloured by specific entropy, show each explosion near the end of its simulated time (20, 22, 28, and 49 ms post-bounce in order of increasing misalignment). Here, it is clear that the dynamics of the most misaligned case (45$\degree$; bottom right) differ significantly from the dynamics of the other three cases. The 15$\degree$ and 30$\degree$ misalignments produce jets that continue to propagate away from the PNS along the poles, as in the aligned case. These jets take slightly longer than the jet in the aligned case to reach a given radius, but the difference in expansion time is relatively small. The case of 45$\degree$ misalignment, however, produces less bipolar outflow geometries. Rather than driving a clear jet, this most misaligned configuration results in cocoon-like bipolar structures surrounded by an equatorial flow. In this paper, we will focus on the nucleosynthesis outputs of these models; the dynamics of the simulations themselves will be discussed in detail in a later paper.

\subsection{Ejecta properties} \label{sec:eject}

\begin{figure}
    \includegraphics[width=\columnwidth]{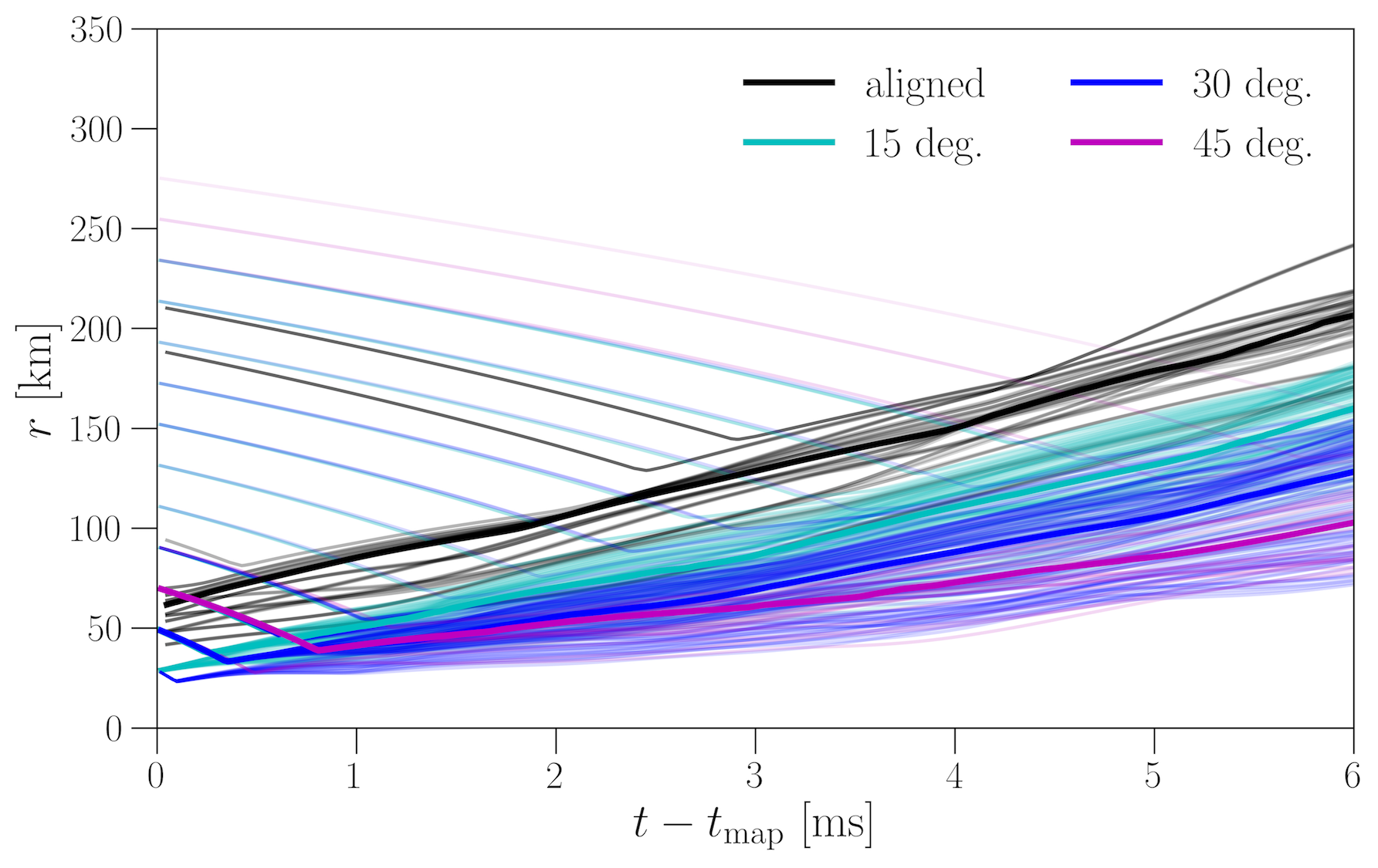}
    \caption{Radial trajectories for all individual ejected tracer particles (light coloured lines) and median trajectories (darker lines) for the four simulations.}
    \label{fig:traj}
\end{figure}

\begin{figure}
    \includegraphics[width=\columnwidth]{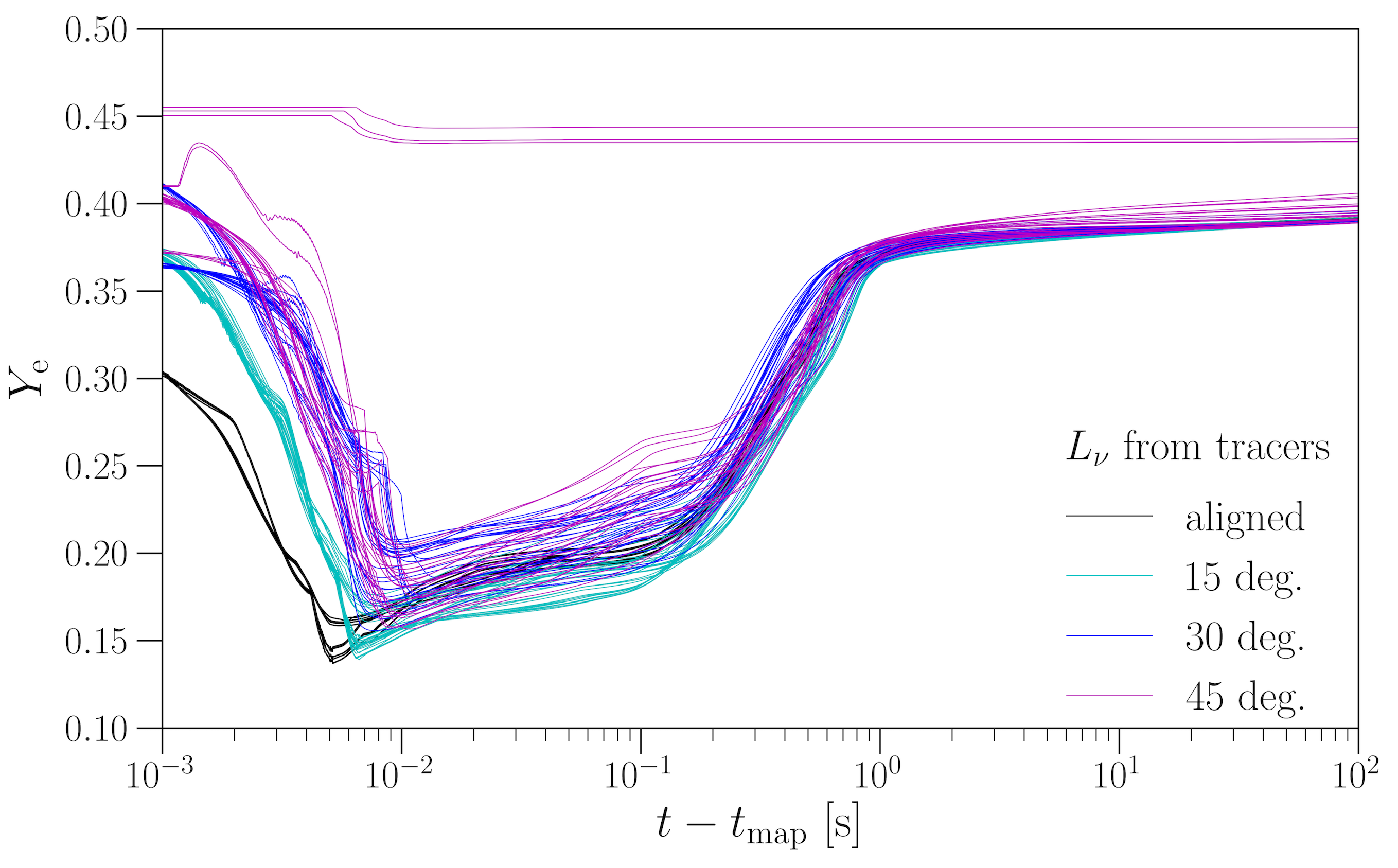}
    \caption{$Y_e$ evolution shown for all tracer particles. Initially, $Y_e$ is directly taken from the tracer particles, but later on $Y_e$ is calculated with \textsc{SkyNet}, where the particles are processed with the neutrino luminosities from the simulations.}
    \label{fig:ye_nulum}
\end{figure}

The trajectories of ejected tracer particles reflect the MHD dynamics of the models and help explain the resulting nucleosynthetic yields. Fig. \ref{fig:tracers3d} shows 3D particle trajectories for representative samples of tracer particles in each of our simulations. Visual comparison between these trajectories and the entropy renderings of Figs. \ref{fig:ent_xz} and \ref{fig:volren} confirms good agreement between the particle dynamics and the MHD dynamics. Although the full 3D trajectories are illustrative, in presenting the analysis we focus on the tracer trajectories along the radial direction. In Fig. \ref{fig:traj}, we show the distance from the PNS center, $r = (x^2 + y^2 + z^2)^{1/2}$, as a function of the time elapsed since mapping tracer particles onto the simulation ($t_\mathrm{map}$) for individual particles, as well as the median of these trajectories for each model. In the aligned case (black), nearly all ejected particles are immediately moving away from the PNS after core bounce. As we go to higher degrees of misalignment, more of the tracer particles instead first move towards the PNS before moving away. This makes sense given the dynamics of the simulations as seen in Figs. \ref{fig:ent_xz} and \ref{fig:volren}. For the more aligned cases, the tracer particles are caught in the powerful jet quickly and pushed away along the $z$-axis, spending less time near the PNS. For models with higher misalignment angles, on the other hand, the tracers can accrete towards the PNS before being dragged away by the slower, less collimated outflows. In all cases, the evolution of $r$ for each particle is dominated by motion along the $z$-axis, since the eventually unbound particles are those which end up in the jets.

There are two competing effects that determine the electron fraction of the material. Densities are higher near the PNS, so tracer particles that get closer in experience higher densities and thus lower $Y_e$. However, if a particle stays near the PNS for too long, it is bombarded by neutrinos, driving $Y_e$ up. For a detailed discussion of the relevant nuclear physics for $r$-process nucleosynthesis and the effects of neutrino interactions, we refer readers to Sec. 3.2 of \citet{moestarobertshalevi2017}. It is clear from Fig. \ref{fig:traj} that eventually ejected tracers in the 30$\degree$ and 45$\degree$ models (blue and magenta, respectively) spend the longest time near the PNS. This trend in particle trajectories is helpful in understanding the corresponding $Y_e$ evolutions, calculated with the neutrino luminosities extracted from the simulations, shown in Fig. \ref{fig:ye_nulum}. The dependence of $Y_e$ on the misalignment angle, as a function of time, is not linear and clear-cut. For example, tracers from the 45$\degree$ model reach low $Y_e$ faster than those from the 30$\degree$ model and reach a minimum value of $Y_e$ comparable to or below that of the more aligned simulations. After this minimum, the tracers from the 45$\degree$ model have more rapidly increasing $Y_e$ than those from the other three models, which can be attributed to the longer time they spend near the PNS experiencing high neutrino flux. The 15$\degree$ model (cyan) has a $Y_e$ evolution nearly identical to that of the fully aligned model (black), but shifted to later times. This delayed but similar evolution is unsurprising given the corresponding particle trajectories as shown in Fig. \ref{fig:traj}; the typical particle ejected in the 15$\degree$ model follows a trajectory almost equivalent to that of the typical particle ejected in the aligned model.

\begin{figure}
    \includegraphics[width=\columnwidth]{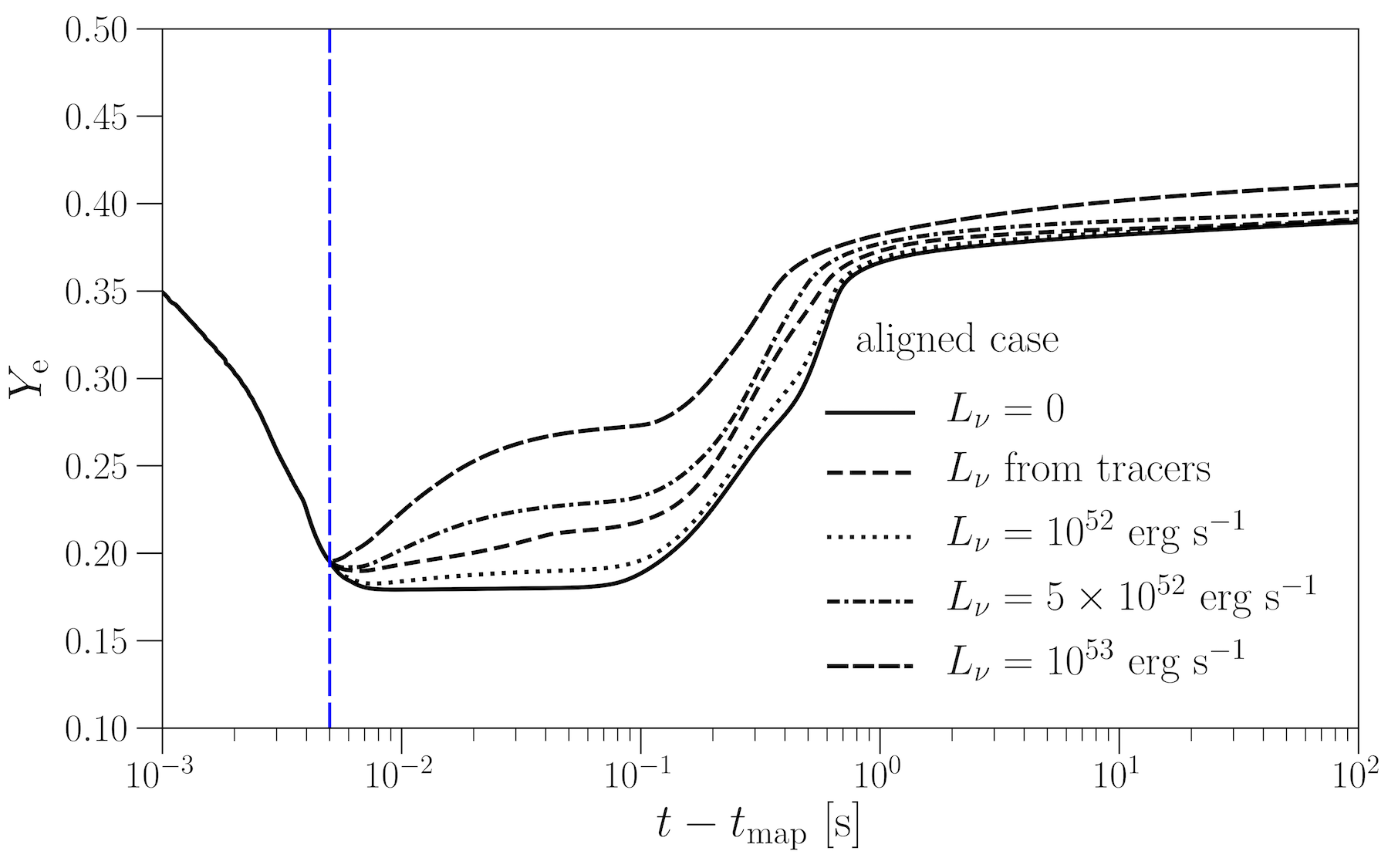}
    \caption{Time evolution of electron fraction $Y_e$ for a single representative tracer particle from the fiducial aligned case, shown for different post-processing options: taking neutrino luminosities directly from the simulation, or setting constant neutrino luminosities ranging from $L_\nu = 0$ to $10^{53}$ erg s$^{-1}$. The vertical dashed line at $\approx5$ ms marks the transition between using values of $Y_e$ recorded by the tracers and calculating $Y_e$ with \textsc{SkyNet}.}
    \label{fig:ye_aligned}
\end{figure}

\begin{figure}
    \includegraphics[width=\columnwidth]{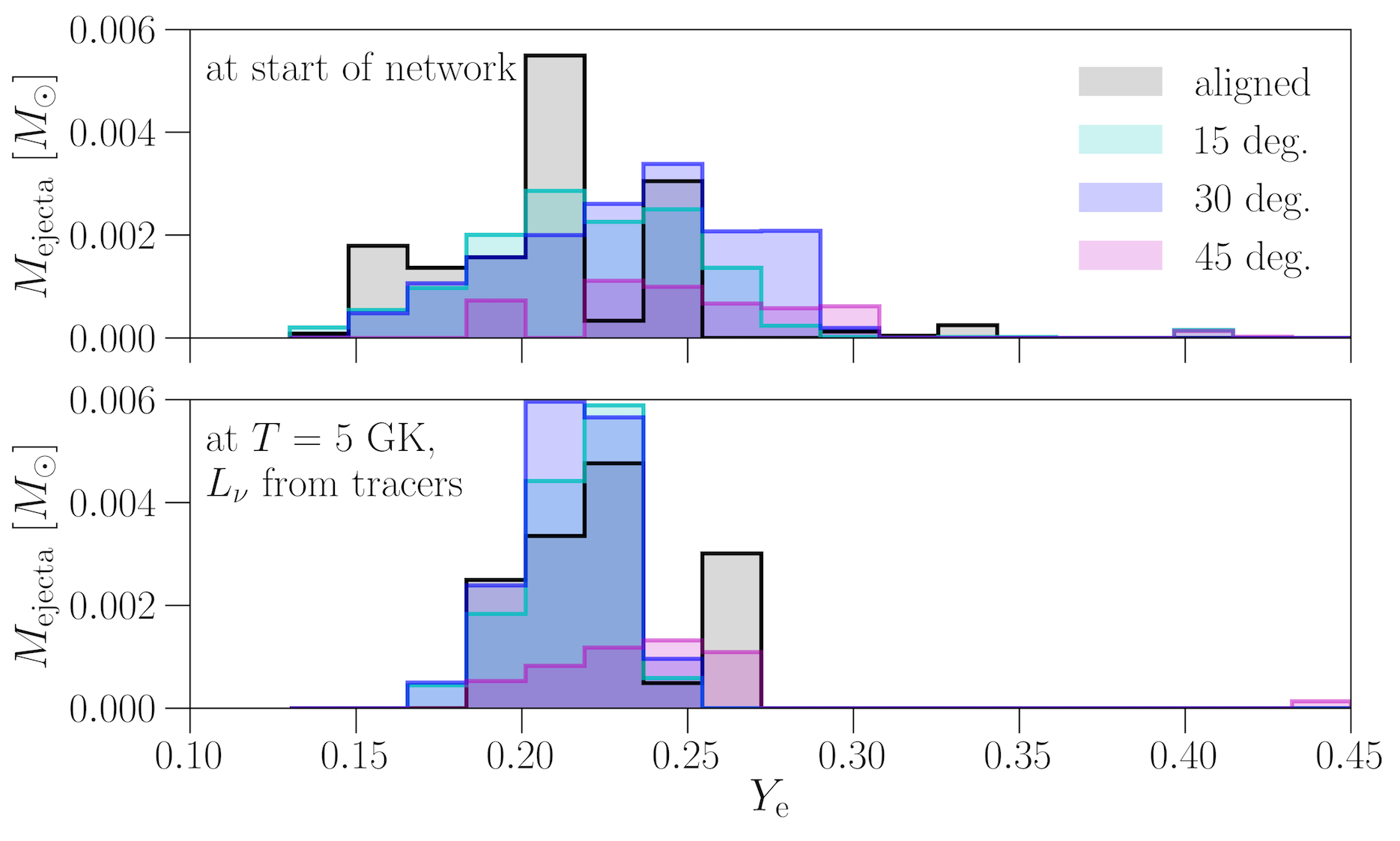}
    \caption{Distributions of $Y_e$ weighted by mass for the four models. In the top panel, the distributions are shown at the start of the \textsc{SkyNet} calculation and are set entirely by the simulations without being affected by post-processing. The bottom panel shows the distributions at a later time, when the temperature has just dropped below the characteristic temperature at which $r$-process begins, $T=5~\mathrm{GK}$. We obtain these distributions by evolving $Y_e$ using \textsc{SkyNet} with the neutrino luminosities set by the simulations.}
    \label{fig:ye_weight}
\end{figure}

As mentioned before, neutrinos radiated by the PNS affect the surrounding material through interactions that convert free neutrons to protons, resulting in higher electron fractions. To check the sensitivity of our models to the effects of neutrinos, we post-process with four different constant neutrino luminosity values. The $Y_e$ evolutions in the case of the aligned model for these values of $L_\nu$, along with those extracted from the simulations, are shown in Fig. \ref{fig:ye_aligned}. Clearly, higher values of $L_\nu$ contribute to an overall increase in $Y_e$. In particular, the effect of the neutrino flux is to reduce the time that the tracer particles spend at relatively low $Y_e$. We find that this trend holds for the misaligned models as well.

The distribution in values of $Y_e$ at the beginning of running \textsc{SkyNet}, weighted by tracer particle mass, is shown for each of the four simulations in the top panel of Fig. \ref{fig:ye_weight}. For the aligned case, the distribution peaks at a low value of $Y_e \gtrsim 0.2$ and is skewed to low $Y_e$. For the case of 15$\degree$ misalignment, the distribution is more symmetric and there is significantly more material with $Y_e > 0.22$ than for the fully aligned case. This trend continues for greater misalignments. The distribution is strongly skewed towards higher $Y_e$ and peaks at $Y_e \approx 0.24$ for 30$\degree$ misalignment, with still much more material with $Y_e > 0.25$. For the most misaligned case of 45$\degree$, the total ejecta mass is low compared to the other simulations (see Table \ref{tab:ejecta}) and it is distributed nearly evenly at higher electron fractions with $0.18 \lesssim Y_e \lesssim 0.31$.

\subsection{Nucleosynthesis yields} \label{sec:comp}

\begin{table}
    \caption{Total and $r$-process ($120\leq A \leq 249$) ejecta masses for the four simulations. The $r$-process abundance yield ($M_{\mathrm{ej,r}}$) depends on the neutrino luminosity set in post-processing, so we show this quantity for every setting: four constant values of $L_\nu$ (in erg s$^{-1}$) and the values taken from the simulation. Note that direct comparisons between the aligned case here and model B13 of \citet{moestarobertshalevi2017} yield inconsistencies due to differences in the initialization of tracer particles.}
    \label{tab:ejecta}
    \begin{tabularx}{\columnwidth}{clcccc}
        \hline \hline
         \textbf{model} &  & \textbf{0 deg.} & \textbf{15 deg.} & \textbf{30 deg.} & \textbf{45 deg.} \\\hline
         $\mathbf{M_\mathrm{ej,tot}}$ &  & 1.41 & 1.32 & 1.55 & 0.51 \\ 
         $[10^{-2} M_\odot]$ &  &  &  &  &  \\ \hline
         $\mathbf{M_{\mathrm{ej,r}}}$ & $L_\nu = 0$ & 1.29 & 1.29 & 1.53 & 0.49 \\

         $[10^{-2} M_\odot]$ & $L_\nu = 10^{52}$ & 1.24 & 1.26 & 1.49 & 0.47 \\
         & $L_\nu$ from sim. & 1.11 & 1.19 & 1.43 & 0.37 \\
         & $L_\nu = 5 \times 10^{52}$ & 0.81 & 0.73 & 0.48 & 0.03 \\
         & $L_\nu = 10^{53}$ & 0.32 & 0.18 & 0.06 & $6 \times 10^{-5}$ \\
         \hline \hline 
    \end{tabularx}
\end{table}

The amount of $r$-process ejecta (defined here as all ejecta with mass number $120\leq A \leq 249$) produced is summarized for each simulation and each choice of post-processing neutrino luminosity $L_\nu$ in Table \ref{tab:ejecta}.

\begin{figure}
	\includegraphics[width=\columnwidth]{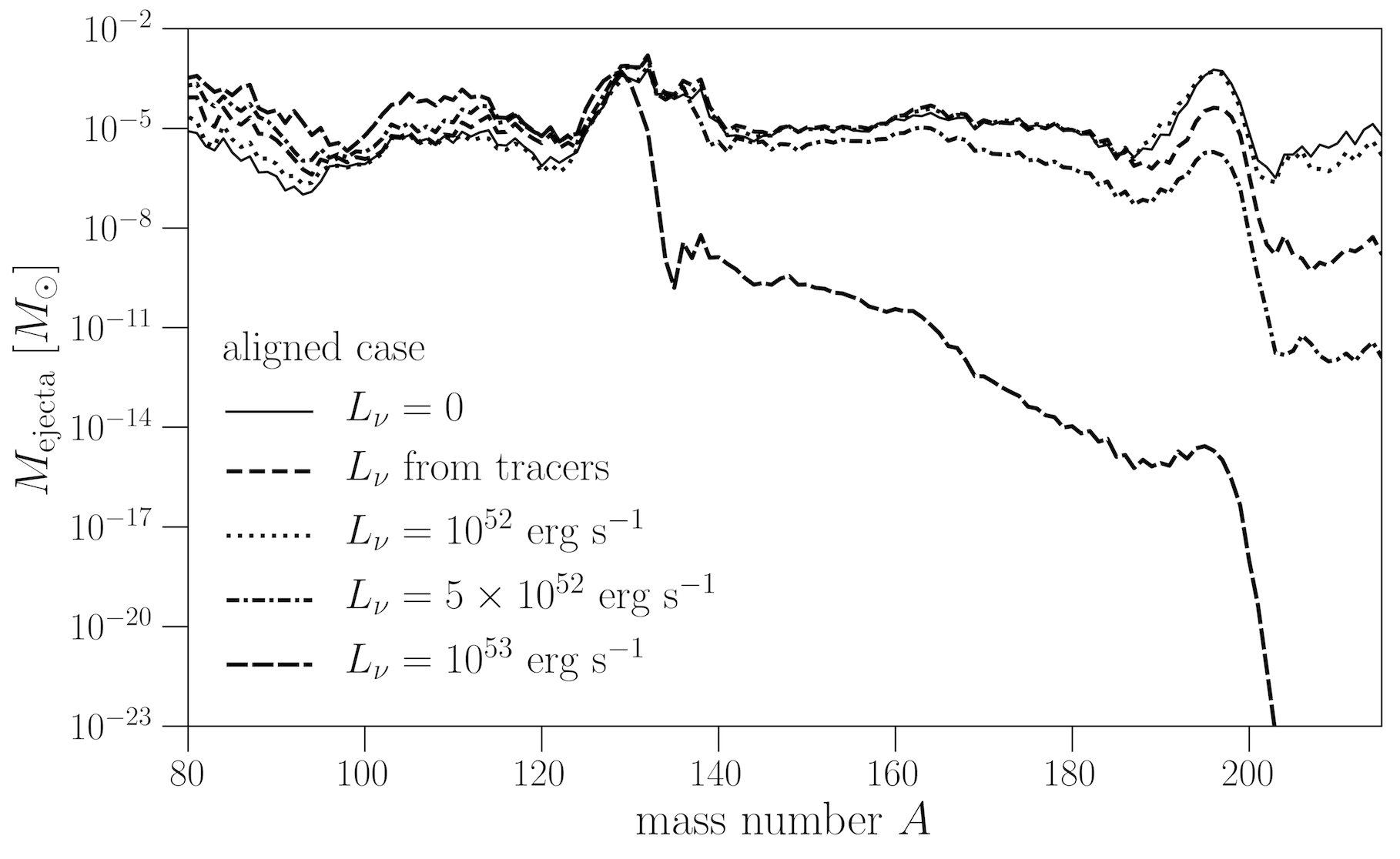}
    \caption{Abundance patterns for the aligned case, as obtained through post-processing the tracer particles. The various curves correspond to the different values used for the neutrino luminosities.}
    \label{fig:abunds_aligned}
\end{figure}

Like the $Y_e$ evolution, the abundance pattern obtained for each model depends on our choice of $L_\nu$, as set in post-processing with \textsc{SkyNet}. This dependence is illustrated for the fiducial aligned case in Fig. \ref{fig:abunds_aligned}, in which we plot the ejecta mass as a function of mass number $A$ for the five different choices of neutrino luminosities. The trends shown for this simulation hold for the other three: for the heaviest nuclei ($A \gtrsim 200$), higher values of $L_\nu$ result in lower abundances, while the opposite is true for light nuclei $A \lesssim 120$. In particular, the third $r$-process peak is weaker for higher neutrino luminosities, and this dependence is strong. For the aligned case, the abundance of third peak elements decreases by $\sim 2-3$ orders of magnitude as we go from $L_\nu = 0$ to $L_\nu = 5 \times 10^{52}~\mathrm{erg}~\mathrm{s}^{-1}$, and drops by another $\sim 8-9$ orders of magnitude at the extreme value of $L_\nu = 10^{53}~\mathrm{erg}~\mathrm{s}^{-1}$. The second peak is far less sensitive to the chosen neutrino luminosity as long as it is below a threshold value of $L_\nu \approx 5 \times 10^{52}~\mathrm{erg}~\mathrm{s}^{-1}$. For reasonable values of $L_\nu$, the second peak is robustly produced in the aligned case. Comparing the trends in abundance patterns to the trends in $Y_e$ produced by varying $L_\nu$, shown in Figs. \ref{fig:abunds_aligned} and \ref{fig:ye_aligned} respectively, provides an intuitive explanation for the reduced production of elements at and beyond the second $r$-process peak for high values of $L_\nu$. As detailed in Sec. \ref{sec:eject}, higher values of $L_\nu$ correspond to less time spent at low $Y_e$. This results in less time for $r$-process nucleosynthesis, which prefers material with low electron fractions, to take place and thus an overall decrease in the abundances of heavy $r$-process elements.

\begin{figure}
	\includegraphics[width=\columnwidth]{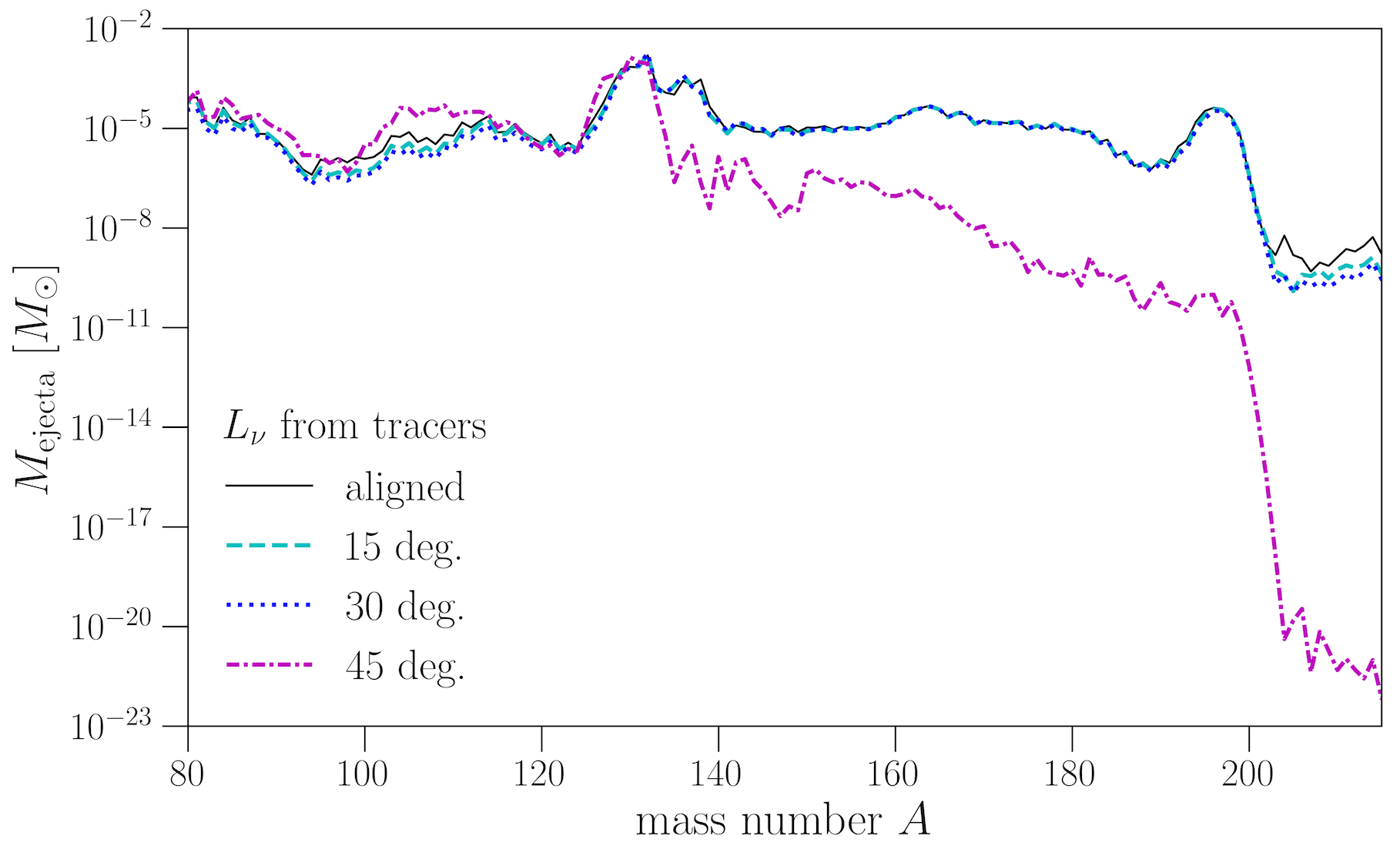}
    \caption{Abundances for the four different models, all obtained using the neutrino luminosities extracted from the simulations. It is clear that the 45$\degree$ misalignment model differs significantly from the other three, particularly in its ability to produce heavy $r$-process elements.}
    \label{fig:abunds_nulum}
\end{figure}

In Fig. \ref{fig:abunds_nulum}, we show the abundance pattern, post-processed with the neutrino luminosities extracted from the tracers, for each of the four simulations. This allows for a direct comparison to determine the effect of misalignment on $r$-process nucleosynthesis yields. Both the second and third $r$-process peaks are quite insensitive to misalignments of 15$\degree$ or 30$\degree$ between the magnetic and rotation axes, but show decreases in abundances for the case of 45$\degree$ misalignment. In this most misaligned model, the third peak is not produced at all when we use the neutrino luminosities from the tracer particles. To understand why the $r$-process abundances depend on the misalignment angle in this way, we examine the $Y_e$ distributions at the time when $r$-process begins, at a temperature of $T \simeq 5\times 10^{9}~\mathrm{K}$. The $Y_e$ distributions at the onset of $r$-process nucleosynthesis ultimately determine the $r$-process element production. These distributions, as evolved by \textsc{SkyNet} with $L_\nu$ taken from the tracers, shown in the bottom panel of Fig. \ref{fig:ye_weight}, differ significantly from the corresponding distributions at the start of the network calculation (top panel). After the network has run to the point when the material has cooled enough for $r$-process to begin, the distributions in $Y_e$ are far more similar for all cases except the 45$\degree$ misalignment than they were at the start of the calculation. All three distributions peak at $Y_e \sim 0.21-0.23$ and have similar amounts of material in each bin of $Y_e$, except for a secondary peak at $Y_e \sim 0.26$ for the aligned case. The 45$\degree$ case, however, clearly differs from the other three in its corresponding distribution of $Y_e$. It peaks at higher $Y_e \sim 0.24-0.25$ and generally has much less material at low $Y_e$ than the more aligned models at the time when $r$-process nucleosynthesis begins.

\section{Discussion}
\label{sec:disc}

The four simulations we have carried out differ only in the axis along which the poloidal magnetic field lies in the initialized core. Our fiducial model is the one in which the magnetic and rotation axes are aligned, while the other three models represent various levels of misalignment: 15$\degree$, 30$\degree$, and 45$\degree$.

Our results suggest several clear trends with the degree of misalignment, particularly in the explosion dynamics, and more complex dependence in the properties of the ejecta and resulting nucleosynthesis yields. In general, higher misalignment results in slower expansion along the polar (rotation) axis and more material along the perpendicular, equatorial axes (Figs.\ref{fig:ent_xz} and \ref{fig:volren}). In other words, more aligned initial fields produce more jetted, bipolar explosions. The ejected material robustly reaches larger distances from the PNS for smaller misalignments (see Fig. \ref{fig:traj}), which is unsurprising given these differences in the dynamics. The distance from the PNS reached by ejected material at a given time after core bounce varies approximately linearly with the misalignment angle.

The effect of misalignment on the $r$-process abundance patterns at a set value of $L_\nu$ is less intuitive. Interestingly, the abundance patterns produced by the 15$\degree$ and 30$\degree$ misaligned models are very nearly identical to that of the aligned model. However, the abundances of second peak and beyond $r$-process elements are highly reduced for the 45$\degree$ model. We explicitly show a comparison of abundance patterns from the four simulations for the case of the neutrino luminosities extracted by the tracer particles (Fig. \ref{fig:abunds_nulum}). In  order to isolate the factors driving this effect, we have also investigated whether it holds for all constant, fixed values of $L_\nu$. We find that as we increase $L_\nu$, the differences between the abundance patterns are magnified. When the neutrinos are effectively turned off ($L_\nu = 0$), all four simulations, including the 45$\degree$ misaligned model, produce nearly identical abundance patterns. As we increase $L_\nu$, we see the abundance pattern of the 45$\degree$ case increasingly deviate from the rest of the models. This indicates that there is a purely dynamical difference between the most misaligned model and the other three models. The effect of this difference in the dynamics on the resultant nucleosynthetic signatures becomes more clear as neutrinos become more important. This can be understood when we consider that slightly different trajectories experience different durations of time near the PNS under neutrino bombardment. The ejecta properties for cases in which tracers dwell near the PNS for longer differ from cases with tracers spending less time near the PNS more significantly when the luminosity of the neutrinos the material interacts with is greater. The effects of slight variations in dynamics, which set particle trajectories, on resultant ejecta properties are thus enhanced for higher neutrino luminosities.

We post-process the results of each simulation with \textsc{SkyNet} using neutrino luminosities from the tracer particles. In addition, we set $L_\nu$ to constant values of $0$, $10^{52}$, $5 \times 10^{52}$, and $10^{53}$ $\mathrm{erg}~\mathrm{s}^{-1}$. We explore this parameter space because the values of $L_\nu$ from our simulations are subject to uncertainties due to imperfections in our neutrino transport treatment -- we use a leakage scheme which, although it has been shown to be a quite good one \citep[e.g.][]{2010CQGra..27k4103O}, is still an approximation. For all four models, the abundance pattern produced for the case in which $L_\nu$ is set by the simulation itself lies somewhere between the abundance patterns of the $L_\nu = 10^{52}$ and $L_\nu = 5 \times 10^{52}~\mathrm{erg}~\mathrm{s}^{-1}$ cases. The same is true for the evolution in $Y_e$. These observations point to a range of interest in $L_\nu$ relevant for making actual predictions of the $r$-process nucleosynthesis yields for our models. The uncertainty in the values of $L_\nu$ extracted by the tracers mainly gives rise to uncertainty in the abundance of third peak elements. By considering the bracketing cases of constant $L_\nu$, $10^{52}$ and $5 \times 10^{52}$ $\mathrm{erg}~\mathrm{s}^{-1}$, we estimate this uncertainty as a factor of 100-1000.

\section{Conclusions}
\label{sec:conc}

We have investigated $r$-process nucleosynthesis in a set of 3D simulations of MR CCSNe, the explosions resulting from highly magnetized, rapidly rotating stellar cores. The four models we simulate represent varying degrees of misalignment between the magnetic and rotation axes of the initialized cores. We run one fully aligned simulation along with three misaligned simulations corresponding to misalignments of 15$\degree$, 30$\degree$, and 45$\degree$. Our aligned model is identical to model B13 of \citet{moestarobertshalevi2017} and similar in dynamics to that of \citet{2012ApJ...750L..22W}. All of our models are initialized with a modified poloidal magnetic field of strength $10^{13}~\mathrm{G}$ and differ only in the angle between the magnetic and rotation axes. Although there are significant uncertainties in the actual magnetic field geometries (and strengths) of MR CCSNe progenitor cores, this work is the first to explore the effect of different field configurations on $r$-process nucleosynthesis in these events. We post-process the simulation data with the nuclear reaction network \textsc{SkyNet} and include weak interactions in order to account for interactions between the ejected material and neutrinos emitted from the cooling PNS. In post-processing, we specify both constant, representative neutrino luminosities and those extracted directly from the tracer particles in our simulations. This allows us to parameterically explore the effect of neutrinos on the $r$-process yields. We summarize our main results below.
\begin{itemize}
	\item A misalignment angle of 45$\degree$ yields an explosion which differs significantly from those of more mild misalignments in terms of both its dynamics and its nucleosynthetic signature.
	\item Higher neutrino luminosities generally result in less $r$-process production, especially for luminosities beyond $L_\nu \sim 5 \times 10^{52}$ erg s$^{-1}$, which respresents the upper bound on realistic values.
	\item An uncertainty of roughly a factor of 2 in the neutrino luminosities from our simulations mainly propagates as an uncertainty in the abundance of third peak $r$-process elements, which we estimate as a factor of 100-1000.
	\item We find robust $r$-process nucleosynthesis for simulations with misalignments of up to 30$\degree$ processed with realistic values for the neutrino luminosities \citep[for further discussion on `realistic' neutrino luminosities, see][]{moestarobertshalevi2017}.
	\item Higher neutrino luminosities accentuate differences in the MHD dynamics produced by distinct misalignments; the $r$-process abundance patterns of the four simulations are nearly identical for $L_\nu = 0$ but diverge with increasing $L_\nu$, with more misaligned models yielding reduced abundances of elements beyond the second peak.
\end{itemize}
We provide video visualizations of our four simulations, viewed as volume renderings coloured by specific entropy, at \url{http://www.astro.princeton.edu/~ghalevi/mrsne_movies}.

\section*{Acknowledgements}
The authors acknowledge J. Lippuner, S. Couch, and N. Choksi for helpful comments on this manuscript, J. Lippuner and L. Roberts for assistance with \textsc{SkyNet}, and K. Parfrey for productive discussions regarding misalignment implementation. PM acknowledges support by NASA through Einstein Fellowship grant PF5-160140. This work used the Extreme Science and Engineering Discovery Environment \citep[XSEDE; ][]{6866038}, which is supported by National Science Foundation grant number ACI-1548562, through allocation TG-AST160049. The authors acknowledge the Texas Advanced Computing Center (TACC) at The University of Texas at Austin (http://www.tacc.utexas.edu). This research also used resources of the National Energy Research Scientific Computing Center (NERSC), a DOE Office of Science User Facility supported by the Office of Science of the U.S. Department of Energy under Contract No. DE-AC02-05CH11231.



\bibliographystyle{mnras}
\bibliography{paper} 




\bsp
\label{lastpage}
\end{document}